\documentclass[journal]{new-aiaa}
\usepackage[utf8]{inputenc}
\usepackage{textcomp}

\usepackage{graphicx}
\usepackage{grffile}
\usepackage{subcaption}
\usepackage{amsmath}
\usepackage[version=4]{mhchem}
\usepackage{siunitx}
\usepackage{longtable,tabularx}
\usepackage{physics}
\usepackage[dvipsnames]{xcolor}

\usepackage{hyperref}
\usepackage[ruled,vlined,linesnumbered]{algorithm2e}
\usepackage[capitalise]{cleveref}

\newcommand\blfootnote[1]{%
	\begingroup
	\renewcommand\thefootnote{}\footnote{#1}%
	\addtocounter{footnote}{-1}%
	\endgroup
}

\newcommand{\ashape}{$\alpha$-shape }

\title{Propagation and reconstruction of re-entry uncertainties using continuity equation and simplicial interpolation\blfootnote{Presented as paper AAS 19-409 at the 29th AAS/AIAA Space Flight Mechanics Meeting, Ka’anapali, HI, 13-17 January 2019}}

\author{Mirko Trisolini \footnote{Postdoc Research Fellow, Department of Aerospace Science and Technology, via La Masa 34, 20156, Milan, Italy.}}
\affil{Polytechnic of Milan, Milan, 20133, Italy}
\author{Camilla Colombo \footnote{Associate Professor, Department of Aerospace Science and Technology, via La Masa 34, 20156, Milan, Italy.}}
\affil{Polytechnic of Milan, Milan, 20133, Italy}

\begin{document}
\sisetup{quotient-mode=fraction}

\maketitle

% ========================================================================
% ABSTRACT
% ========================================================================

\begin{abstract}
This work proposes a continuum-based approach for the propagation of uncertainties in the initial conditions and parameters for the analysis and prediction of spacecraft re-entries. Using the continuity equation together with the re-entry dynamics, the joint probability distribution of the uncertainties is propagated in time for specific sampled points. At each time instant, the joint probability distribution function is then reconstructed from the scattered data using a gradient-enhanced linear interpolation based on a simplicial representation of the state space. Uncertainties in the initial conditions at re-entry and in the ballistic coefficient for three representative test cases are considered: a three-state and a six-state steep Earth re-entry and a six-state unguided lifting entry at Mars. The paper shows the comparison of the proposed method with Monte Carlo based techniques in terms of quality of the obtained marginal distributions and runtime as a function of the number of samples used.
\end{abstract}

% ========================================================================
% INTRODUCTION
% ========================================================================

\section{Introduction}  \label{sec:intro}
\lettrine{T}{he} study and prediction of re-entry trajectories is a challenging and complex task. Several factors may influence the accuracy of these predictions, such as the knowledge of the initial entry conditions, the ballistic coefficient of the satellite, and the density of the atmosphere. The uncertainties associated with these parameters can affect the evolution of the re-entry trajectory, and influence the prediction of several quantities of interest such as the impact location, and the mechanical and thermal loads on the spacecraft. It is thus important to include statistical verification inside the mission design process, considering the effects of uncertainties in assessing possible off-nominal scenarios. Combining this analysis with operational constraints and safety margins can ultimately improve the safety and robustness of mission designs. This type of statistical analysis can be applied to several aspects of the re-entry problem. The design of crewed vehicles and their navigation algorithms can strongly benefit from an uncertainty analysis given the strict requirements on the sustainable loads and on the landing location \citep{hoelscher2007orion,rea2007comparison,putnam2008improving,brunner2008skip,putnam2014feasibility}. The design of exploration probes and sample return missions can also gain from uncertainty quantification to improve the robustness of the mission \citep{spencer1996mars,striepe2006mars,nelessen2019mars} with the aim to increase the landing mass and precision and to provide more robust designs for the thermal protection and parachute deployment systems. Finally, even the destructive re-entry of satellites and the prediction of their re-entry footprint, in particular for uncontrolled re-entries can benefit from uncertainty assessment. Until 2013, there has been an average of 93 uncontrolled re-entries per year \citep{Pardini2013} and with the increasing space activities of the past few years, the frequency of re-entering objects is destined to grow rapidly. Despite these objects usually pose a marginal risk for people on the ground, it is still necessary to verify their compliance with the casualty risk regulations \citep{OConnor2008}. This ultimately means studying the break-up of the spacecraft and the demise of its parts \citep{dobarco2005orsat, Koppenwallner2005, Gelhaus2014}. Considering the effects of uncertainties, it is possible to assess the demise probability of specific components \citep{TLC18Sensitivity} and the statistical distribution of the casualty area of surviving components.

The traditional procedure to assess uncertainties is based on a Monte Carlo (MC) dispersion analysis, where, through a large number of simulations over randomly sampled initial conditions and parameters, the joint Probability Density Function (PDF) is estimated through a frequentist approach. This type of simulation provides a reliable way to estimate the evolution of uncertainties given their straight-forward implementation and they can effectively captures nonlinearities in the system. However, in general, to obtain convergent statistics, the number of MC simulations must increase. Consequently, for multi-dimensional state spaces and nonlinear dynamics, such as the ones associated with re-entry scenarios, they can become expensive \citep{niederreiter1992random, luo2017review}. Nonetheless, many space missions’ uncertainty analyses have been carried out using a Monte Carlo approach. Examples are the mission design of the Mars Pathfinder \citep{spencer1996mars}, of the Mars Science Laboratory \citep{striepe2006mars} missions, and of the future Mars2020 exploration mission \citep{nelessen2019mars}. Simulation software such as the NASA Dynamics Simulator for Entry, Descent, and Surface Landing (DESENDS) \citep{cameron2016dsends} uses a Monte Carlo-based dispersion analysis. In addition, MC simulations have been extensively used for the design verification of entry guidance algorithms \citep{hanson2004test,rea2007comparison,putnam2008improving,lu2008predictor,lu2017verification}. In these works, the performance, robustness, and reliability of the algorithm is tested considering a dispersion in the initial conditions and parameters of the re-entry. The results of the MC simulations in terms of mean and standard deviation are compared to the target for the landing accuracy, and the number of occurrences for violation of specified safety thresholds is considered. Also destructive re-entry codes such as ESA DRAMA \citep{Gelhaus2014} perform re-entry analyses using an MC approach for the design and verification of satellite compliance to the casualty risk regulations.

Instead of MC simulation, several other techniques have been applied to the propagation of uncertainties \citep{luo2017review}. For example, Unscented Transformation (UT), where only a few, deterministically chosen, samples are propagated (the sigma points). From these samples is then possible to obtain a second order approximation of the first two moments of the probability distribution. Other methodologies are instead based on the creation of a surrogate model that can more efficiently estimate the uncertainty when compared to MC \citep{mehta2017sensitivity,sanson2019space,dell2015probabilistic}. An example of such methodologies is Polynomial Chaos Expansion (PCE) \citep{prabhakar2010polynomial,jones2013satellite}, where the inputs and outputs of a system are represented via series approximation. They provide an efficient way to propagate uncertainties building an explicit functional representation of the output uncertainty with respect to the inputs. These techniques are computationally efficient, even in high dimensions, even though they rely on the estimation of the uncertainties through a frequentist approach as MC does. Another common uncertainty propagation technique is based on Differential Algebra (DA) \citep{lunghi2018atmospheric}, which allows the computation of arbitrary order expansion with respect to the initial condition. DA can be used in conjunction with MC simulations, substituting them with Taylor expansion. Gaussian Mixture Models (GMM) \citep{luo2017review} can also be used to propagate uncertainties. In this method, the nonlinearity and non-Gaussian behavior of the probability density function is captured by using multiple Gaussian distributions. However, the effects of nonlinearities have to be considered by possibly splitting and merging the different components of the mixture.

The approach we propose uses the continuity equation to directly propagate the probability density along with the dynamics of the system, thus obtaining a systematic evolution of the probability density. This methodology has been applied in the study of the dynamical evolution and formation of stellar systems, planetary ring structures, and interplanetary dust \citep{Chandrasekhar1943,Gorkavyi1997,McInnes1995}. Additionally, it has been used for the propagation of space debris fragments following a break up of satellites in orbit \citep{Letizia2018,FCL19AAS} and for the propagation of uncertainties in planetary re-entries \citep{Halder2011,TC19UncertainityAAS,TC2020ReentryStarling1,LTFC20AsteroidReentry}. Such a methodology is opposed to MC simulations where the distribution is approximated through many realizations: while with MC methods we propagate individual realization of the initial PDF, with a continuum-based approach we propagate the ensemble of realizations. When not only the first moments of the distribution are needed but its whole shape is of interest, MC simulations require a large number of samples. Instead, the proposed continuum-based propagation method allow the knowledge of the probability density at specific points in the state space can enable a reconstruction of the probability density function with a reduced number of samples.

The challenge of using a method based on the continuity equation is the post-processing of the data. As we are propagating a finite set of initial points, with their probability density, it is then necessary to reconstruct this density in the state space starting from discrete values. In \citep{Halder2011} a reconstruction methodology has been proposed, which replicates the binning process of MC simulations. For each bin, the density is computed as the mean of the density values of the data points contained in it. This method can be used if the PDF data is uniformly distributed in each bin in all dimensions and if the enclosed volume of the scattered data in each bin is equal \citep{Hoogendoorn2018}. However, even if the initial distribution of samples is uniform, its evolution in time does not necessarily remain uniform and the enclosed volume of the data in each bin does not remain equal. Therefore, such a methodology generates results which present poor agreement with corresponding Monte Carlo simulations \citep{Halder2011, Hoogendoorn2018}. In addition, such a methodology still uses a number of samples comparable to Monte Carlo methods. We instead propose to reconstruct the density using a simplex-based linear interpolation methodology \citep{Stoer2002} that uses the concept of \ashape \citep{Edelsbrunner1994}, which adapts to the evolution of the shape of the state space volume. In addition, we increase the accuracy of the linear interpolation by including the derivative information into the linear interpolation scheme using the reduced order dual Taylor expansion \citep{Kraaijpoel2010}.

The paper presents three relevant test cases, which include uncertainties in the initial conditions, in the ballistic coefficient of the satellite, and in the atmospheric density. The first test case considers a strategic re-entry on Earth using a three-state dynamics, the second test cases expands the first one to a more complex dynamics, and the third test case features the lifting re-entry of a probe in Martian atmosphere. The results are presented as one-dimensional and two-dimensional marginal distributions of the relevant parameters. The distribution for derived quantities of interest, such as the heat rate and the dynamic pressure, are also presented. The paper is organized as follows: \cref{sec:propagation} describes the methodology for the propagation of the uncertainties using a continuum-based approach; \cref{sec:reconstruction} describes the methodology used for the reconstruction of the probability density and for the computation of the marginal probabilities. \cref{sec:test_cases} presents the application of the density propagation and reconstruction procedure to relevant test cases. \cref{sec:conclusion} contains the discussion and the conclusions.

% ========================================================================
% UNCERTAINTY PROPAGATION
% ========================================================================

\section{Uncertainty propagation}  \label{sec:propagation}
The proposed methodology uses the continuity equation to propagate the initial joint probability distribution function related to the uncertainties in the initial conditions and parameters and assess its evolution throughout the re-entry process under the influence of the re-entry dynamics. The expression for the continuity equations is as follows \citep{Gorkavyi1997}:

\begin{equation}  \label{eq:continuity}
    \pdv{n(\mathbf{x}, t)}{t} + \nabla  \mathbf{f}(\mathbf{x}) = \dot{n}^{+} - \dot{n}^{-},
\end{equation}

where $\mathbf{x}$ is the vector of the state variables, $n(\mathbf{x}, t)$ is the joint probability distribution function (PDF) at time $t$, $\mathbf{f}(\mathbf{x})$ represents the forces acting on the system and takes into account slow, continuous phenomena such as gravity and atmospheric drag, and $\dot{n}^{+}$ and $\dot{n}^{-}$ represent the fast and discontinuous events (i.e. sources and sinks). For the case in exam, the source and sink terms were neglected. Knowing the initial density distribution $n(\mathbf{x}, 0)$, \cref{eq:continuity} allows for the propagation of the density evolution in time, in a system with the equation of the dynamics. When applied to the propagation of uncertainties, the density represents the probability distribution. This is a Partial Differential Equation (PDE) with the PDF, $n(\mathbf{x}, t)$, being the dependent variable. Such an equation regulates the conservation of the total probability mass of the joint PDF through its spatial-temporal evolution due to the forces acting on the system. \cref{eq:continuity} can be solved using the Method Of the Characteristics (MOC), where the partial differential equation is transformed into a set of Ordinary Differential Equations (ODE). As it is convenient to express the evolution of the re-entry trajectory using parameters such as the altitude, the relative velocity, and the flight path angle, we follow the approach to express the continuity equation in the state space of the problem in exam, writing the divergence in rectangular coordinates \citep{Gorkavyi1997}. In the generic case of $d$ number of variables, \cref{eq:continuity} can be re-written in rectangular coordinates as follows:

\begin{equation}
    \pdv{n}{t} + \pdv{n}{\alpha_{1}} v_{\alpha_{1}} + \ldots + \pdv{n}{\alpha_{d}} v_{\alpha_{d}} + \bigg[ \pdv{v_{\alpha_{1}}}{\alpha_{1}} + \ldots + \pdv{v_{\alpha_{d}}}{\alpha_{d}} \bigg]  n = 0
\end{equation}

where $\alpha_{i}$ are the state variables and $v_{\alpha_{i}}$ the corresponding forces. The sink and source terms have been neglected in this case. Applying the method of characteristics, the PDE can be reduced to the following system of ODEs:

\begin{equation}  \label{eq:char}
    \begin{cases}
     \dv{t}{s} &= 1  \\
     \dv{\alpha_1}{s} &= v_{\alpha_1}(\alpha_1, \ldots, \alpha_d) \\
     &\vdots \\
     \dv{\alpha_d}{s} &= v_{\alpha_d}(\alpha_1, \ldots, \alpha_d) \\
     \dv{n}{s} &= \bigg[ \pdv{v_{\alpha_{1}}}{\alpha_{1}} + \ldots + \pdv{v_{\alpha_{d}}}{\alpha_{d}} \bigg]  n(\alpha_1, \ldots, \alpha_d)
     \end{cases}
\end{equation}

where $s$ is the independent variable. In this study, the continuity equation is applied to two sets of equations of motion. First, a \emph{three-state} representation models, which describes the evolution of the re-entry through radius ($r$), velocity ($v$), and flight-path angle ($\gamma$) under the influence of the planet gravity and the atmospheric drag. The dynamics of this model is described in \cref{eq:eqmot_3state} and considers a planar motion over an non-rotating planet.

\begin{equation} \label{eq:eqmot_3state}
    \begin{cases}
        \dot{r} = v  \sin{\gamma}  \\
        \dot{v} = -\frac{1}{2 \beta}  \rho  v^2 - g  \sin{\gamma} \\
        \dot{\gamma} = \bigg( \frac{v}{r} - \frac{g}{v} \bigg)  \cos{\gamma} + \frac{1}{2 \beta} \frac{C_L}{C_D}  \rho  v %%\\
        %%\dot{\beta} = 0
    \end{cases}
\end{equation}

where $\rho$ is the atmospheric density, $g$ is the gravitational acceleration, $\beta = \frac{ m }{ C_D S }$ is the ballistic coefficient, $R_p$ is the radius of the planet, $C_L$ is the lift coefficient, $C_D$ is the drag coefficient, and $S$ is the object cross-section. Applying the MOC of \cref{eq:char} to this set of equations, the variation in time of the probability density, $n$, is obtained as follows: 

\begin{equation}  \label{eq:ndot_3state}
%\dot{n} = - \bigg[ \frac{v}{\beta}  \rho + \sin{\gamma}  \bigg( \frac{v}{r} - \frac{g(r)}{v} \bigg) \bigg]  n.
\dot{n} = - \bigg[ \pdv{\dot{r}}{r} + \pdv{\dot{v}}{v} + \pdv{\dot{\gamma}}{\gamma} \bigg]  n.
\end{equation}

If uncertainties in additional parameters, other than re-entry states, want to be considered, it is necessary to include them in the equations of motion and add their contribution to the evolution of the density in \cref{eq:ndot_3state}. For example, in the case in exam, we want to consider an uncertainty over the ballistic coefficient, $\beta$, and include the possibility to take into account uncertainty in the atmospheric density (through an atmospheric correction coefficient, $\xi$). The resulting system of equation for the propagation of the characteristics becomes as follows:

\begin{equation} \label{eq:system3state}
\begin{cases}
\dot{r} = v  \sin{\gamma}  \\
\dot{v} = -\frac{1}{2 \beta}  \rho(r, \xi)  v^2 - g(r)  \sin{\gamma} \\
\dot{\gamma} = \bigg( \frac{v}{r} - \frac{ g(r) }{ v } \bigg)  \cos{\gamma} + \frac{1}{2 \beta} \frac{C_L}{C_D}  \rho(r, \xi)  v \\
\dot{\beta} = 0 \\
\dot{\xi} = 0 \\
\dot{n} = - \bigg[ \frac{v}{\beta}  \rho(r, \xi) + \sin{\gamma}  \bigg( \frac{v}{r} - \frac{g(r)}{v} \bigg) \bigg]  n.
\end{cases}
\end{equation}

Therefore, the result is the augmented state, ($r$, $v$, $\gamma$, $\beta$, $\xi$), to be propagated. %For the description of the atmosphere and of the gravity field, different models can be used, provided they can be described by differentiable functions.
The second model considered in this study is a \emph{six-state} representation, which describes the three-dimensional translational re-entry over a rotating Earth. For the case in exam, we decided to express the equations of motion using the radius as the independent variable, instead of the time (de facto obtaining a \emph{five-state} model). On one side, this allows showing the flexibility of the continuum propagation and, on the other, it simplifies the representation of the output of the propagation as, for example, it is more convenient to extract information at the landing instant, which corresponds to the final radius in the propagation. The set of equations of motion is the following:

\begin{equation} \label{eq:system6state}
\begin{cases}

\dv{\lambda}{r} = \frac{ \sin (\chi ) }{ r \cos (\varphi ) \tan (\gamma ) } \\

\dv{\varphi}{r} = \frac{ \cos (\chi )}{r \tan (\gamma )} \\

\dv{v}{r} = - \frac{ v \rho(r, \xi) }{ 2 \beta \sin(\gamma) } - \frac{1}{v}  \bigg( g_r(r, \varphi) + \frac{ \cos(\chi)
	g_\varphi( r, \varphi) }{\tan(\gamma ) } \bigg) + \frac{ r \omega_p^2 \cos(\varphi) }{ v }  \bigg( \cos(\varphi) - \frac{\cos(\chi) \sin(\varphi) }{ \tan(\gamma) } \bigg) \\

\dv{\gamma}{r}  = \frac{ \alpha \rho(r, \xi) }{ 2 \sin(\gamma) } - \frac{1}{v^2 \tan(\gamma) }  \bigg( g_r(r, \varphi) + \cos(\chi) \tan(\gamma) g_\varphi(r,\varphi) - \frac{ v^2 }{ r } \bigg) + \frac{ 2 \omega \sin(\chi) \cos(\varphi) }{ v \sin(\gamma) } + \\ \quad\quad + \frac{ r \omega_p^2 \cos(\varphi) }{ v }  \bigg( \frac{ \cos(\varphi) }{ \tan(\gamma) } + \cos(\chi) \sin(\varphi) \bigg) \\

\dv{\chi}{r}  = \frac{ \sin(\chi) \tan(\varphi) }{ r \tan(\gamma)} + \frac{ 2 \omega_p }{ v }  \bigg( \frac{ \sin(\varphi) }{ \sin(\gamma) } - \frac{ \cos(\chi) \cos(\varphi) }{ \cos(\gamma) } \bigg) + \frac{ \sin(\chi) }{ v^2 \sin(\gamma) \cos(\gamma) }  \bigg(  r \omega_p^2 \sin(\varphi) \cos(\varphi) - g_\varphi(r, \varphi) \bigg) \\

\dv{\beta}{r} = 0 \\

\dv{\xi}{r} = 0 \\

\dv{n}{r} = - \bigg[ \pdv{\lambda^\prime}{\lambda} + \pdv{\varphi^\prime}{\varphi} + \pdv{v^\prime}{v} + \pdv{\gamma^\prime}{\gamma} + \pdv{\chi^\prime}{\chi} + \pdv{\beta^\prime}{\beta} + \pdv{\xi^\prime}{\xi} \bigg]  n

\end{cases}
\end{equation}

where $\lambda$ is the longitude, $\varphi$ the latitude, $v$ the velocity, $\gamma$ the flight-path angle, $\beta = \frac{ m }{ C_D S }$ the ballistic coefficient, $\alpha = \frac{ C_L S }{ m }$ a modified lift coefficient, $\xi$ an atmospheric correction coefficient, and $g_r$ and $g_\varphi$ are the radial and transversal components of the gravitational acceleration, respectively. The expression for the derivative of the density as a function of $r$ was not expanded for a better readability. In this expression, the primes refers to derivatives with respect to the radius. As it is possible to observe, even in this second re-entry model it has been included the possibility to consider uncertainties in the ballistic coefficient and in the atmospheric density, by including the coefficient $\beta$ and $\xi$ into the extended state of the problem.

\cref{eq:system3state,eq:system6state} must be integrated numerically and the integration can be performed using a standard ODE solver such as Runge-Kutta. In this way, the time evolution of the density in the state space can be obtained as a function of the considered independent variable. As the solution for the re-entry problem is not analytical, it is necessary to sample the uncertainty distribution in the initial states and to propagate the trajectory and the probability density for each sample point.

% ========================================================================
% DENSITY RECONSTRUCTION
% ========================================================================

\section{Interpolation method}  \label{sec:reconstruction}
Once the sampled initial conditions have been propagated using \cref{eq:char}, it is necessary to reconstruct the probability density in the domain at each time step to obtain the total uncertainty and the marginal distributions. The proposed approach is integrated with the continuum propagation described in \cref{sec:propagation} to reconstruct the probability density from few samples, leveraging on the knowledge of the value of the density that is propagated alongside the characteristics. As it can be observed in \cref{fig:strategic_scatter}, during the evolution of the re-entry trajectory, not only the density but also the state-space volume changes and deforms. This is why the method proposed in \citep{Halder2011}, which performs a uniform binning to estimate the probability density by averaging the its value in each bin, provides inaccurate results when the state space starts to deform. In our methodology, instead, we follow the variation of the state space volume by creating a simplicial (i.e. a generalization of the notion of triangle or tetrahedron to arbitrary dimensions) representation and interpolating the scattered data. Scattered data interpolation is a complex task, and is even more challenging when the considered data has more than three dimensions as it is for the case in exam. Several techniques exist to interpolate scattered data in 1D and 2D such as spline interpolation \citep{awanou2005multivariate}, multi-variate polynomial \citep{alfeld1989scattered,barthelmann2000high}, and radial basis function \citep{buhmann2003radial}. However, these techniques can be difficult to extend to arbitrary dimensions, or require regular grid. In this work, we propose to reconstruct the density using linear interpolation based on the Delaunay triangulation \citep{Preparata1985} of the sampled points. We thus seek to approximate a multivariate function $f: \; \mathbf{X} \rightarrow \mathbb{R}$ with $\mathbf{X} \subset \mathbb{R}^d$, where the elements of $\mathbf{X}$ are denoted by $\mathbf{x} \equiv (x_1, x_2, \ldots, x_d)$. We propose a simplicial-based interpolation given its possibility to be extended to arbitrary dimensions and its capability of allowing for the direct inclusion of derivative information to improve its accuracy \citep{Kraaijpoel2010}. Additionally, this methodology preserves the values at the nodes of the triangulation. This is important as it allows the conservation of a crucial information that is the value of the probability density at the sampled points as provided by the continuum-based propagation.

This section is structured as follows: \cref{subsec:linear_int} describes the linear interpolation methodology; \cref{subsec:alpha} discusses the limitations of the Delaunay triangulation and proposes a different approach using $\alpha$-shapes; \cref{subsec:dual-taylor} proposes an improved interpolation methodology; \cref{subsec:marginals} describes the procedure adopted to integrate the probability density and obtain the marginal distributions.

%--------------------------------------------------------------
% LINEAR INTERPOLATION
%--------------------------------------------------------------
\subsection{Linear interpolation}  \label{subsec:linear_int}
As previously mentioned, the linear interpolation we adopt is based on a simplicial representation obtained through a Delaunay triangulation, which is unique for a given set of points. Given its heritage, uniqueness, and capability to be extended to arbitrary dimensions, it has been selected in this work to construct the simplical complex (i.e. the union of the simplices forming the triangulated state space) from the propagated scatter data. The construction of the Delaunay triangulation has been carried out using the \emph{python} package \emph{Delaunay} of the \emph{scipy} \citep{scipy} library. Each of the data points in the considered state space (e.g. $r$, $v$, $\gamma$, $\beta$) represents the coordinates of a vertex $V$ of the simplicial complex. At the same time, the probability densities relative to each vertex are the data values of the function $f$ to be interpolated. In general, the simplicial-based linear interpolation can be expressed as follows:

\begin{equation}  \label{eq:simp_int}
    \mathbf{L}(\mathbf{x}) = \sum_{i=1}^{N} \lambda_i(\mathbf{x}) f_i,
\end{equation}

where $N$ is the number of vertices of the simplex (i.e. its dimensionality), $\mathbf{L}(\mathbf{x})$ is the linear interpolation at a point $\mathbf{x}$ inside the considered simplex, $\lambda_i(\mathbf{x})$ is the $i$-th barycentric coordinate of the point $\mathbf{x}$, and $f_i$ is the value of the function (the density in our case) at the $i$-th vertex. Once the barycentric coordinates of the vertices of the simplex are found, the linear interpolation of \cref{eq:simp_int} can be performed. In other words, the linear interpolation is obtained through a weighted average of the value of the function at the simplex vertices, with the weights being the barycentric coordinates of the vertices (i.e. the distances of the vertices form the barycenter of the $d$-dimensional simplex).

%--------------------------------------------------------------
% ALPHA SHAPE
%--------------------------------------------------------------

\subsection{Alpha shapes}  \label{subsec:alpha}
\begin{figure}[hbt!]
    \centering
    \begin{subfigure}[b]{0.45\textwidth}
        \includegraphics[width=\textwidth]{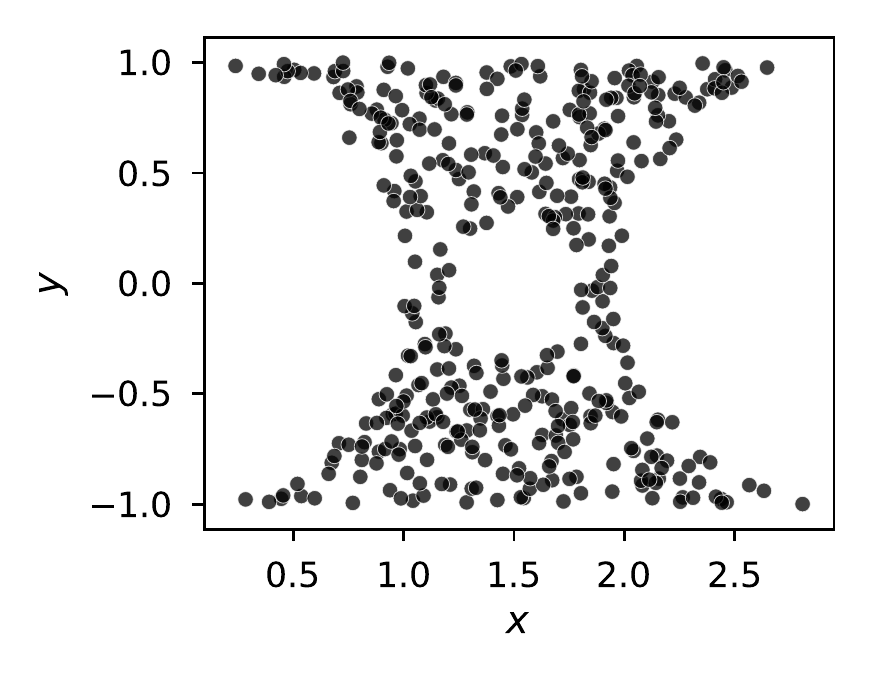}
        \caption{ }
        \label{fig:scatter_test}
    \end{subfigure}
    ~
    \begin{subfigure}[b]{0.45\textwidth}
        \includegraphics[width=\textwidth]{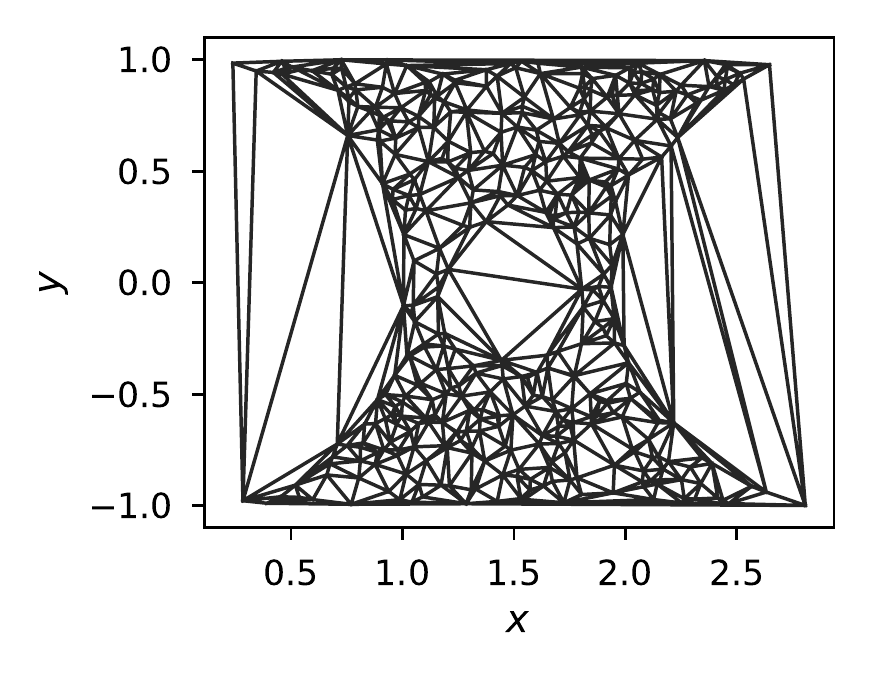}
        \caption{ }
        \label{fig:convex_hull}
    \end{subfigure}

    \caption{Example of a concave set of points (a) and of its Delaunay triangulation (b).}
\end{figure}

A drawback of Delaunay triangulation is that it is based on the convex-hull of a set of points. Therefore, if the reconstructed shape is concave additional unwanted triangles will be generated. For example, \cref{fig:scatter_test} shows a concave set of test points, which also contains a hole. It can be observed from \cref{fig:convex_hull} that the Delaunay triangulation generates simplices for the entire convex-hull and fills the hole inside the set of points with additional simplices, which can compromise the accuracy of the interpolation.

Therefore, we introduce the concept of $\alpha$-shape \citep{Edelsbrunner1994}. The $\alpha$-shape,  $\mathcal{C}_\alpha (V)$ of a set of points $V \subset \mathbb{R}^d$ is a subset of the Delaunay triangulation, $\mathcal{DT}(V)$. The objective of the \ashape is to eliminate from the Delaunay triangulation all the excess simplices that are formed when a shape is not convex. As these simplices tend to be elongate, it is possible to use a test based on the radius of the circum-hyper-sphere (the equivalent of the circum-radius in $d$ dimensions), $\sigma_T$, to prune triangles that are deemed too elongated by setting a threshold on the radius of the circum-hyper-sphere. Specifically, a simplex ($\Delta_T$) belonging to the Delaunay triangulation also belongs to the \ashape if 

\begin{itemize}
	\item[i.] $\sigma_T < \alpha$ and the hyper-sphere of radius $\sigma_T$ is empty, or
	\item[ii.] $\Delta_T$ is a face of another simplex in $\mathcal{C}_\alpha (V)$
\end{itemize}

where $\alpha$ is a hyper-parameter, which has to be selected by the user. We present a possible way to select $\alpha$ in \cref{subsubsec:cval}. The first fo the presented condition is referred to as the \emph{alpha test}. Following this definition and exploiting the properties of the Delaunay triangulation it is possible to build the \ashape through the following steps as described by Edelsbrunner \citep{Edelsbrunner1994}:

\begin{itemize}
    \item[1.] Compute the Delaunay triangulation of $V$, knowing that the boundary of the $\alpha$-shape is contained in it;
    \item[2.] Determine $\mathcal{C}_\alpha (V)$ by inspecting all the simplices $\Delta_T \in \mathcal{DT}(V)$: if the circum-hyper-sphere $\sigma_T$ is empty and $\sigma_T < \alpha$ we accept $\Delta_T$ as a member of $\mathcal{C}_\alpha (V)$, together with all its faces;
    \item[3.] All $d$-simplices of $\mathcal{C}_\alpha (V)$ make up the interior of the $\alpha$-shape.
\end{itemize}

Given the aforementioned procedure, the $\alpha$-shape can be constructed starting from the Delaunay triangulation and removing all the simplices that do not pass the \emph{alpha test}. As for a $d$-simplex belonging to a Delaunay triangulation the circum-hyper-sphere is empty by definition \citep{Preparata1985}, the \emph{alpha test} indicates that the simplex belongs to $\mathcal{C}_\alpha (V)$ if $\sigma_T < \alpha$. Therefore, the test only requires the computation of the radius of the  circum-hyper-sphere for a generic $d$-dimensional simplex that is \citep{coxeter1973regular}

\begin{equation} \label{eq:cradius}
    \sigma_T = R = \sqrt{- \frac{\mathbf{CM}^{-1}_{11}}{2}}
\end{equation}

\begin{figure}[hbt!]
    \centering
    \includegraphics[width=.5\textwidth]{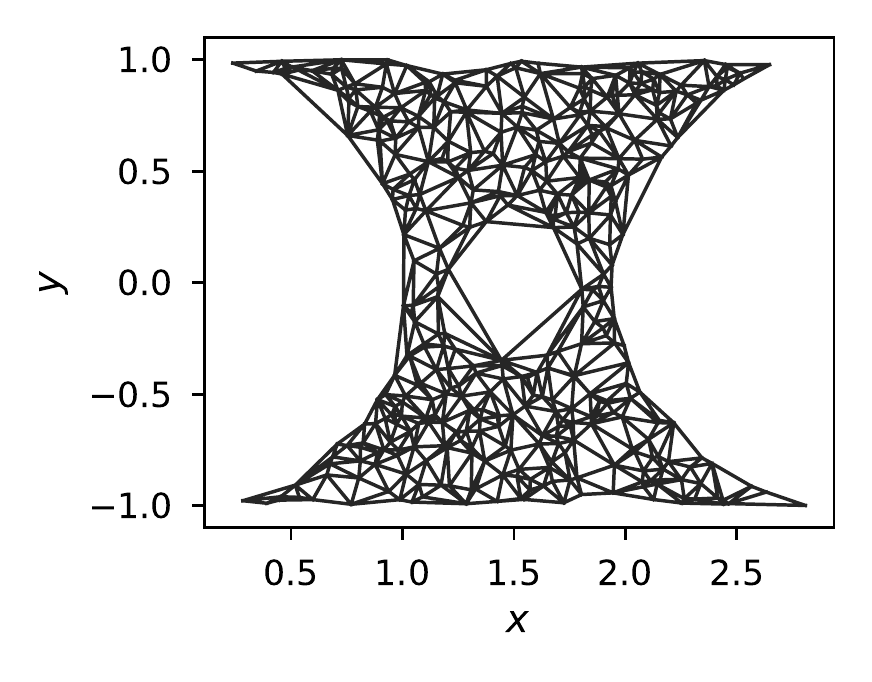}
    \caption{Example of $\boldsymbol{\alpha}$-shape for the set of test points.}
    \label{fig:alpha_hull}
\end{figure}

Applying the $\alpha$-shape algorithm to the set of points of \cref{fig:scatter_test} with an $\alpha$ value of 0.25 we get the $\alpha$-shape of \cref{fig:alpha_hull}, thus improving the shape reconstruction with respect to the Delaunay triangulation of \cref{fig:convex_hull}.

The interpolation using the Delaunay triangulation of \cref{eq:simp_int} can be directly extended to the $\alpha$-shape, the only difference being that the simplices used are the one belonging to the \ashape and not to the Delaunay triangulation. To give an example of the linear interpolation using $\alpha$-shape, it is possible to associate a weight to each point of \cref{fig:scatter_test}. For example we can assume that

\begin{equation}  \label{eq:ftest}
    f(\mathbf{x}_i) = y_i^2.
\end{equation}

\begin{figure}[hbt!]
    \centering
    \begin{subfigure}[b]{0.48\textwidth}
        \includegraphics[width=\textwidth]{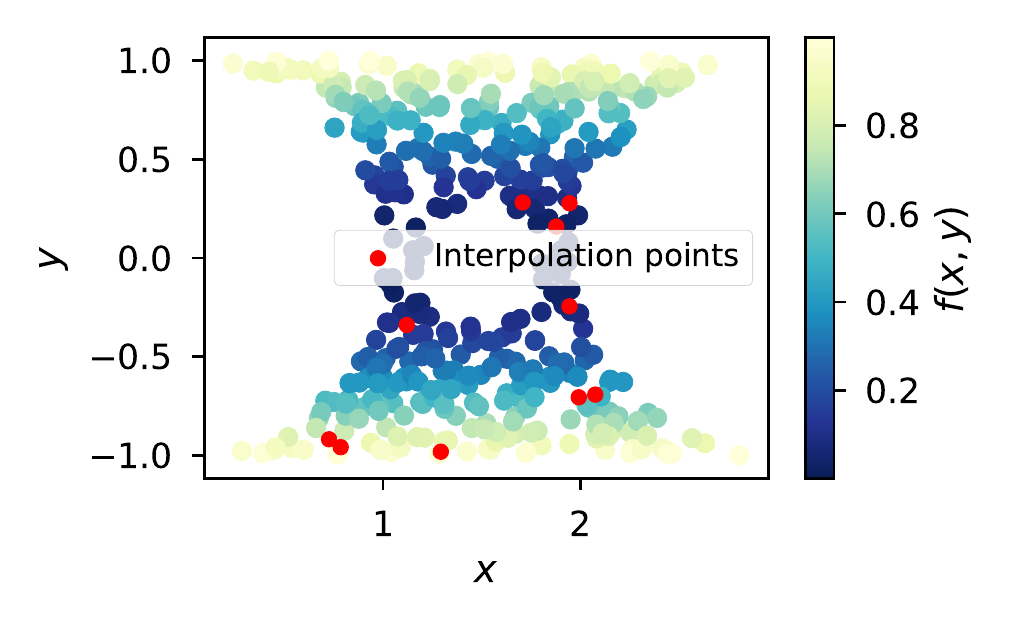}
        \caption{ }
        \label{fig:alpha_test}
    \end{subfigure}
    ~
    \begin{subfigure}[b]{0.48\textwidth}
        \includegraphics[width=\textwidth]{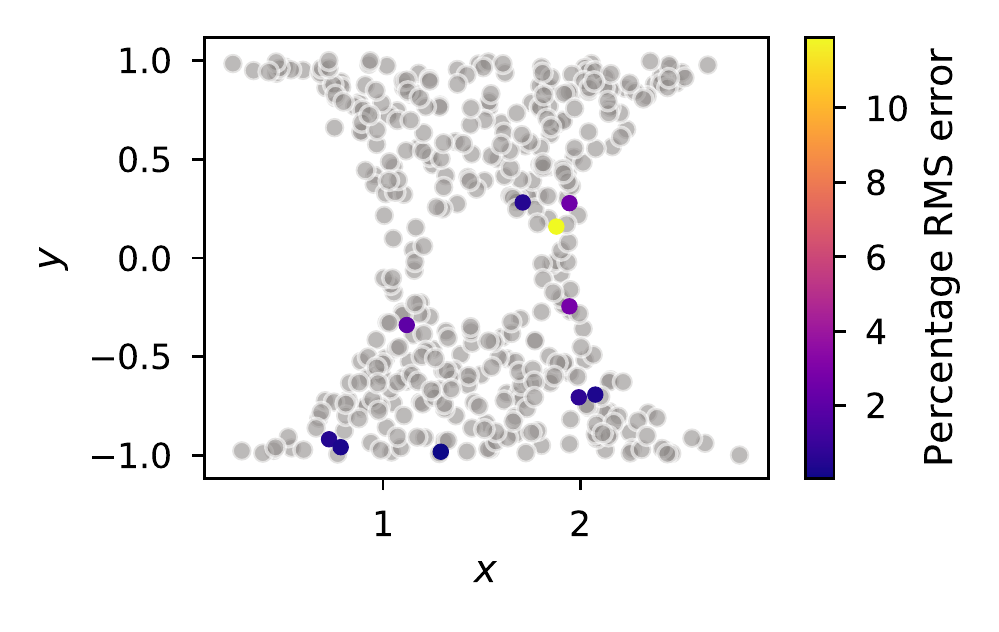}
        \caption{ }
        \label{fig:rms_error}
    \end{subfigure}

    \caption{Interpolated function and selected interpolation points (a) and RMS percent error at the interpolation points (b).}
\end{figure}

To test the interpolation technique, we randomly select ten points from the set, we remove them and we generate the $\alpha$-shape, we then interpolate the function at the coordinates of the ten test points (\cref{fig:alpha_test}). Finally, we check the root mean square (RMS) error between the interpolated values and the actual value of the function at these test points. \cref{fig:rms_error} shows the results of this test for the ten test points, highlighting the RMS percent error. The average error is 2.17\%, while the maximum is 11.89\%.

% ..............................................................
% ALPHA VALUE ESTIMATION
% ..............................................................

\subsubsection{Alpha value estimation} \label{subsubsec:cval}
The value of $\alpha$ in the construction of the \ashape is a hyper-parameter, which determines the simplices that will be removed from the initial Delaunay triangulation and, therefore, how closely the \ashape will resemble the shape of the actual state space (\cref{subsec:alpha}). To determine the $\alpha$ value, it is possible to perform a $k$-fold cross-validation, combined with a stochastic search algorithm. Specifically, we use a Differential Evolution (DE) search method \citep{Storn1997} as it proved effective in the optimization of tuning parameters and hyper-parameters \citep{schmidt2019performance}. In the $k$-fold cross-validation, the available set of points is subdivided into $k$ folds. One of the $k$ folds is then held out and used as validation set, while the other $k-1$ folds are combined into a training set. In this case, the points of the $k-1$ folds are used to construct the $\alpha$-shape, then the interpolation at the points of the remaining fold is carried out. As we are interested in providing the best shape approximation for our data-set, we consider two conditions:

\begin{enumerate}
	\item[1.] if any of the test points lies outside the \ashape constructed with the training data the value of $\alpha$ is rejected;
	\item[2.] if all the test points are within the $\alpha$-shape, we compute the overall volume as the sum of the volumes of the simplices.
\end{enumerate}

We repeat this procedure holding out a different fold each time. The output is then the average of the scores obtained for each fold. Throughout this study, the selected number of folds has been $K=5$. The result of the $k$-fold cross-validation depends on the value of $\alpha$; we apply a DE algorithm to find the value of $\alpha$ that minimizes the volume of the \ashape computed following the previous points. Specifically, we use the \emph{differential\textunderscore evolution} function of the Python \emph{scipy.optimize} library \citep{scipy}. Differential evolution is a stochastic direct search method in which a vector of parameters of size $N_p$, the population size, is evolved for a specified number of generations $N_g$ to find a global optimal solution. The initial population is randomly chosen inside the provided bounds and a uniform distribution for the parameters in the search space is assumed. In our case, this translates into selecting a value of $\alpha_i$ inside the interval $(\alpha_{min}, \alpha_{max})$. The new parameters vectors are generated using the \emph{best1bin} strategy \citep{mezura2008multi}. For this study the mutation rate has been set to $\eta_m = 0.5$ and the recombination rate to $\eta_r = 0.7$, which are both typical values. The population size has been set to $N_p = 40$ and the number of generations to $N_g = 60$. The presented strategy, however, can become computationally cumbersome when the dimension of the problem and/or the number of points increases. For example, to obtain the alpha values for the test case of \cref{subsec:test_strategic} takes around 60 seconds on the laptop used for the simulations in the study. This computational time can be too long with respect to the other operations performed. Possible mitigation strategies are the use of the same $\alpha$ value for different snapshots as long as the state space does not deforms considerably, the use of the $\alpha$ value of the previous snapshot as the starting point for the current one to improve the convergence, a larger parallelization, and a reduction in the $k$-fold cross-validation.
For the cases in which this procedure is deemed impractical, alternative procedures can be used to select the $\alpha$ value. For example, the distances between each point and its nearest neighbor can be evaluated. Then, the value of $\alpha$ can be selected as the average, maximum, minimum, or median value among these distances \citep{teichmann1998surface}. For example, the minimum value will tend to discard more elongated triangles, but can also exclude parts of the tails of the distribution. The selection among this value can be performed by testing them on selected points, similarly to the cross-validation previously described. In addition, they can be used as starting values for the differential evolution algorithm.

%--------------------------------------------------------------
% DUAL TAYLOR EXPANSION
%--------------------------------------------------------------

\subsection{Gradient enhanced interpolation: the reduced dual Taylor expansion}  \label{subsec:dual-taylor}
As shown in \cref{fig:rms_error}, a direct application of the linear interpolation methodology can result in relative errors larger than 10\% for selected points. We can expect that with an increase in the number of dimensions of the problem and complexity of the state space also the accuracy of the interpolation may degrade. An error in the interpolation directly maps into an error in the computation of the integral of the probability density and therefore worsen the approximation of the marginals, especially if large errors concentrate in high-density areas. Therefore, we propose a novel methodology that integrates supplementary derivative data into the presented interpolation (\cref{eq:simp_int}) scheme to improve the accuracy of the approximation. The method we propose replaces the value of the function at the simplex vertices $f_i$ with its \emph{reduced dual Taylor expansion} \citep{Kraaijpoel2010}. The procedure applies to all schemes that are based on function values at discrete nodes, provided the derivative data is available at the same nodes as the function values. This characteristic perfectly fits the interpolation procedure we propose, allowing for its extension and enhancement. The $n$-th order \emph{reduced dual Taylor expansion} of the $m$-th kind $\mathcal{D}_x^{mn}$ is defined as \citep{Kraaijpoel2010}:

\begin{equation}  \label{eq:red-dual-taylor}
\mathcal{D}_x^{mn}[f] := \sum_{|\kappa| \leq n} {\frac{1}{\kappa!} C_{|\kappa|}^{mn} (x - .)^\kappa f^{(\kappa)} (.) },
\end{equation}

with:

\begin{equation}
C_{|\kappa|}^{mn} := \begin{pmatrix}
m + n \\ m
\end{pmatrix}^{-1} \begin{pmatrix}
m + n - |\kappa| \\ m
\end{pmatrix}.
\end{equation}

where $C_{|\kappa|}^{mn}$ are the reduction coefficients. The only difference between a standard dual Taylor expansion and \cref{eq:red-dual-taylor} are exactly these coefficients. The modified expression can be used to raise the order of generic multivariate approximation schemes, provided they use a polynomial approximation of data at specified nodes and that the derivative information is available at these nodes.The implementation of \cref{eq:red-dual-taylor} in our linear approximation scheme is straightforward: we just need to replace the input samples of $f$ with the corresponding samples of $\mathcal{D}_x^{mn}[f]$, which can be obtained using the supplementary derivative data. The interpolation scheme of \cref{eq:simp_int} is then replaced by:

\begin{equation}  \label{eq:taylor_int}
    \mathbf{\widetilde{L}}(\mathbf{x}) = \sum_{i=1}^{N} \lambda_i(\mathbf{x}) \mathcal{D}_x^{mn}[f](\mathbf{x}_k).
\end{equation}

The approximation order will then be raised from $m$ to $m + n$ \citep{Kraaijpoel2010}. The presented procedure perfectly adapts to the chosen methodology for the propagation of uncertainties (\cref{sec:propagation}). In fact, as we already compute the Jacobian of the dynamics, it is of little effort and computational cost to include the expressions for the derivative of the density with respect to the integration variables and return it as an additional output of the density propagation. \cref{eq:gradient} provides an example for the dynamics of \cref{eq:system3state}.

\begin{equation}  \label{eq:gradient}
    \begin{cases}
        \pdv{n}{r} &= -\bigg( \frac{v \sin{\gamma}}{r^2} - \pdv{g(r)}{r} \; \frac{\cos{\gamma}}{v} - \pdv{\rho(r, \xi)}{r} \; \frac{v}{\beta} \bigg) \; n  \\
        \pdv{n}{v} &= -\bigg( \frac{ g(r) \cos{\gamma} }{ v^2 } -           \frac{\sin{\gamma}}{r} - \frac{\rho(r, \xi)}{\beta} \bigg) \; n \\
        \pdv{n}{\gamma} &= -\bigg( \frac{ g(r) \sin{\gamma} }{ v } - \frac{ v \cos{\gamma} }{ r } \bigg) \; n \\
        \pdv{n}{\beta} &= - \frac{v \rho(r, \xi)}{\beta^2} \; n \\
        \pdv{n}{\xi} &= -\frac{v}{\beta} \pdv{\rho(r, \xi)}{\xi} \; n
    \end{cases}
\end{equation}

With this procedure we obtain at each node the required information for a direct application of \cref{eq:taylor_int}. To demonstrate the improvement introduced by the gradient-enhanced interpolation, the test case of \cref{fig:alpha_test} is performed again, this time using the \emph{reduced dual Taylor expansion}. For the function of \cref{eq:ftest}, the derivatives with respect to $x$ and $y$ are:

\begin{equation}  \label{eq:test_grad}
    \begin{cases}
        \pdv{f}{x} &= 0 \\
        \pdv{f}{y} &= 2 \; y.
    \end{cases}
\end{equation}

\begin{figure}[hbt!]
    \centering
    \includegraphics[width=.5\textwidth]{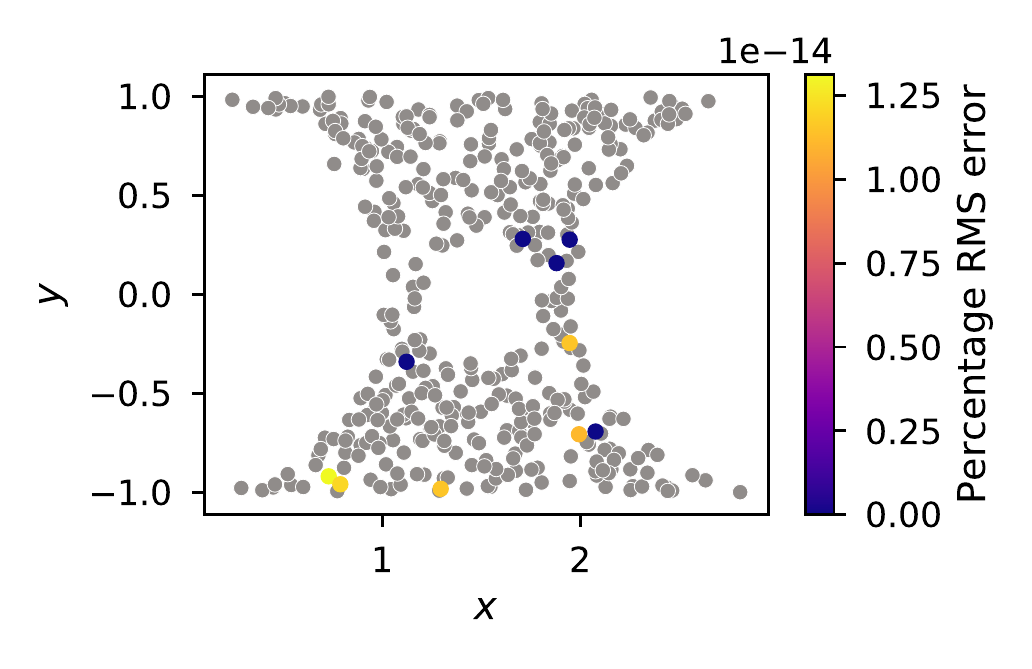}
    \caption{RMS percent error for the interpolation enhanced by the \emph{reduced dual Taylor expansion}.}
    \label{fig:grad_rms_error}
\end{figure}

Including this information, we can perform the interpolation for the same test points. \cref{fig:grad_rms_error} shows the resulting RMS percent error, which is close to zero. The introduction of the derivative information has thus strongly improved the results of the linear interpolation. It has to be noted that this is a particular case: the interpolated function is quadratic (\cref{eq:test_grad}) and with the reduced dual Taylor expansion we introduce a second order approximation, which matches the order of the function to be interpolated.

%--------------------------------------------------------------
% MARGINALIZATION
%--------------------------------------------------------------

\subsection{Marginalization}  \label{subsec:marginals}
The density reconstruction using \emph{alpha shapes} and the \emph{reduced dual Taylor expansion} can be used to obtain the marginal densities of the relevant variables. In our application, this allows for a better understanding of the re-entry under uncertainties, predicting how they influence the evolution in time of the re-entry velocity, altitude, flight-path angle, and derived quantities such the dynamic pressure and heat rate.
Given the generic $d$-dimensional state space $P_s = \{ x_1, x_2, \ldots, x_d \}$, the one-dimensional and two-dimensional marginals can be computed as follows:

\begin{align} 
    m_{x_m} &= \int_{P_s \setminus \{x_m\}} n(x_1, \ldots, x_d) \; dx_1 \ldots dx_{m-1} \, dx_{m+1} \ldots dx_d \label{eq:marginal1} \\
    m_{x_m x_n} &= \iint_{P_s \setminus \{x_m, x_n\}} n(x_1, \ldots, x_d) \; dx_1 \ldots dx_{m-1} \, dx_{m+1} \, dx_{n-1} \, dx_{n+1} \ldots dx_d, \label{eq:marginal2}
\end{align}

which is an integration of the probability density over all the dimensions in the state space, except the ones of the marginal axes. These expressions for the computation of the marginals apply to continuous functions. However, for the case in exam, only a discrete representation of the probability density is available through the interpolation procedure described in the previous sections. The computation of the marginals is here applied to the snapshots obtained through integration of \cref{eq:char}, i.e. the scattered data in the sate space at each time step separately.
% ..............................................................
% INTEGRATION
% ..............................................................
%\subsubsection{Integration}  \label{subsubsec:integration}
%This section describes the procedure adopted to compute the integral of \cref{eq:marginal1,eq:marginal2}.
As the procedure is similar for both one-dimensional and two-dimensional marginals, for a clearer explanation we focus here on the computation of one-dimensional marginals. To do so, it is necessary to adapt \cref{eq:marginal1} to scattered data. In addition, it is necessary to consider the nature of the data of the problem and, specifically, the evolution of the state space geometry with time. As shown for example by \cref{fig:strategic_scatter}, during the re-entry process, the state space tends to suffer from a substantial deformation. We can thus obtain elongated state spaces, which are difficult to interpolate altogether, even when $\alpha$ shapes are employed. In addition, this strategy can reduce memory usage issues that can arise when generating the \ashape in high dimensions. Consequently, it was decided to overcome this issue by using a binning strategy to compute the marginals. First, we select the axis along which the marginal must be computed, $x_m$. Then, this axis is subdivided into $N_b$ equally distributed bins, whose width is given by:

\begin{equation}  \label{eq:binsize}
    s_{b} = \frac{\max(x_m) - \min(x_m)}{N_b}.
\end{equation}

\begin{figure}[hbt!]
    \centering
    \begin{subfigure}[b]{0.45\textwidth}
    \includegraphics[width=\textwidth]{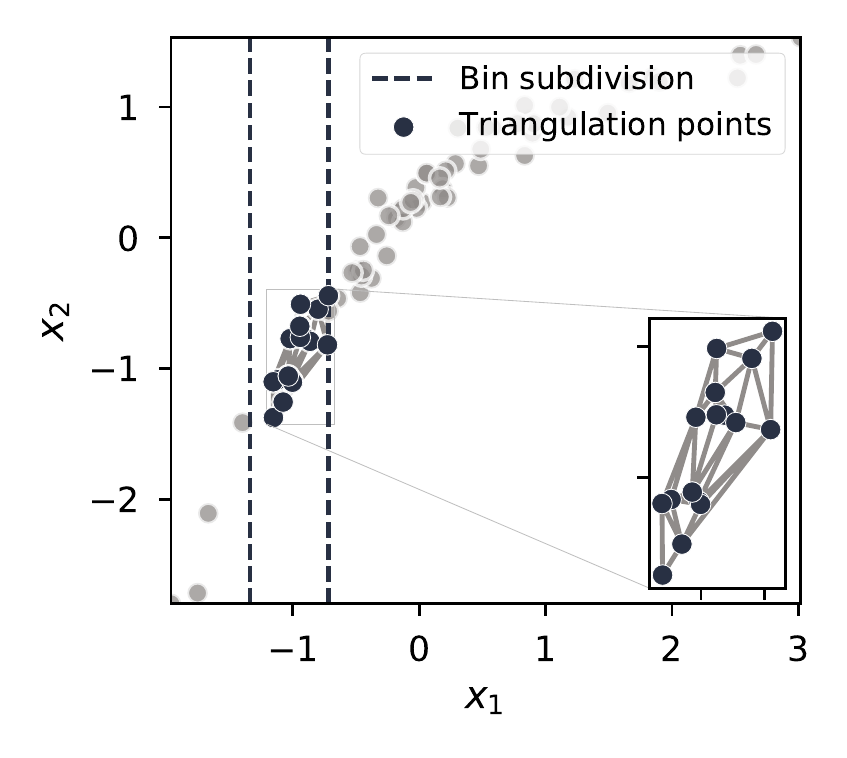}
    \caption{ }
    \label{fig:bin_marginal}
    \end{subfigure}
    ~
    \begin{subfigure}[b]{0.45\textwidth}
    \includegraphics[width=\textwidth]{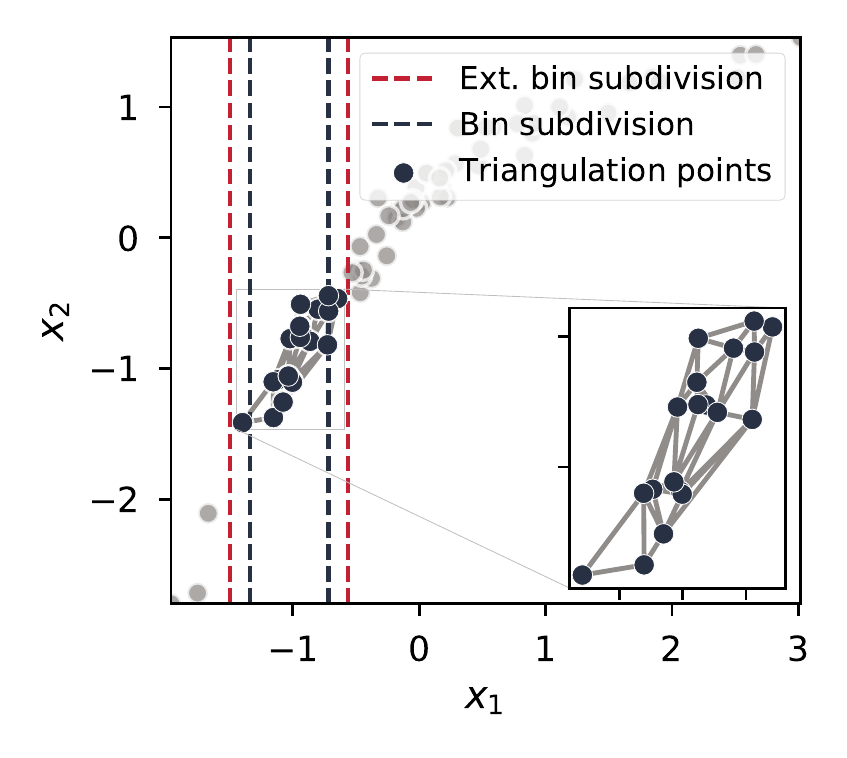}
    \caption{ }
    \label{fig:ext_bin_marginal}
    \end{subfigure}

    \caption{Examples of $\boldsymbol{\alpha}$-shape construction for a bin (a) and for an extended bin (b).}
\end{figure}

In this way, slices of the original set of points are constructed to obtain a set $B$ containing the $N_b$ subsets of the scattered data in which $x_m$ is subdivided:

\begin{equation}  \label{eq:set_of_bins}
B = \big\{ \mathbf{x} \in V \; | \: x_{b, i} \leq x_m \leq x_{b, i+1} \; \forall \: i \in {1, \ldots, N_b} \big\},
\end{equation}

where $x_m$ is the value of the $m$-th coordinate of the point $\mathbf{x}$ and $x_{b, i}$ is the coordinate of the $i$-th bin edge. For each subset of points contained in $B$ an \ashape is constructed to perform the interpolation and compute the integral of the density relative to the considered bin. We then compute the marginal probability for each bin dividing the integral by the width of the bin so that the marginal is given by

\begin{equation}  \label{eq:marginal}
    m_{x_m} = \bigg\{ \frac{W(B_i)}{s_b} \quad \forall \: B_i \in B \bigg\},
\end{equation}

where $W(B_i)$ is the integral of the probability density relative to the $i$-th set of points $B_i$ obtained as expressed by \cref{eq:set_of_bins}. At this point, it is important to outline the computational procedure for this integral. For the sake of clarity, let us consider a two-dimensional example for which \cref{fig:bin_marginal} shows a sample distribution of scattered data with an elongated state space. Suppose we are interested in the marginal distribution relative to $x_1$. For each bin subdivision, as highlighted in \cref{fig:bin_marginal}, an alpha-shape is created using the points inside the bin. However, if we only use the points strictly contained inside the bin, the triangulation will exclude a portion of the volume of the state space, with a consequent underestimation of the integral. Therefore, each bin is expanded using buffer widths on both sides (\cref{fig:ext_bin_marginal}). In this way, the expanded bin, $B_i^\prime$, allows for a triangulation, which fully includes the original bin.

Once this \emph{extended} \ashape has been created, the interpolation is performed. The interpolating points are the barycenters of the simplices of the \ashape so that the integral of the density inside the extended bin is

\begin{equation}  \label{eq:ext_simplex_integral}
    W(B_i^\prime) = \sum_{k=1}^{N_{s_i}} n(\mathbf{x}_{G_k}) \; v_k,
\end{equation}

where $n(\mathbf{x}_{G_k})$ is the interpolated value of the probability density at the barycenter of the $k$-th simplex ${x}_{G_k}$ as obtained using \cref{eq:taylor_int} of the gradient-enhanced interpolation scheme described in \cref{subsec:dual-taylor}, $N_{s_i}$ is the number of simplices contained inside the extended bin $B_i^\prime$, and $v_k$ is the volume of the $k$-th simplex. The integral computed using \cref{eq:ext_simplex_integral} refers to the extended bin. It is thus necessary to re-scale this value. We do not know with precision the volume of the state space contained inside the bin. However, it will be contained inside the volume of the \ashape constructed with the initial bin, $\mathcal{C}_\alpha (B_i)$ (\cref{fig:bin_marginal}), and the one constructed with the extended bin, $\mathcal{C}_\alpha (B_i^\prime)$ (\cref{fig:ext_bin_marginal}). Therefore, the integral is scaled with the average between these two volumes as follows:

\begin{equation}  \label{eq:simplex_integral}
    W(B_i) = W(B_i^\prime) \; \frac{ \text{Mean}(v_{ \mathcal{C}_\alpha (B_i) }, v_{ \mathcal{C}_\alpha (B_i^\prime) }) }{ v_{ \mathcal{C}_\alpha (B_i^\prime) } }.
\end{equation}

Repeating this procedure for each bin in the set $B$ allows for the computation of the one-dimensional marginal. For the computation of the two-dimensional marginals the procedure is analogous, with the binning performed over the two dimensions of the marginals ($x_m$, $x_n$).

%This can be computed for a generic $d$-dimensional simplex using the barycentric transform matrix (\cref{eq:bary_transform_matrix}) as follows \citep{Preparata1985}:
%
%\begin{equation}  \label{eq:simplex_volume}
%    v_k = \frac{ | \det(\mathbf{A}) | }{d!}.
%\end{equation}

% ==============================================================
% TEST CASES
% ==============================================================

\section{Test cases} \label{sec:test_cases}

This section presents a series of relevant test cases to the problem in exam, applying the propagation methodology described in \cref{sec:propagation} and the reconstruction methodology outlined in \cref{sec:reconstruction} to planetary entry cases with different dynamical, gravitational, and atmospheric models. In this way, we show the applicability of the continuum propagation with models of different complexity. The results in terms of marginal distributions are compared with Monte Carlo simulations in terms of accuracy and execution time. In addition, for selected cases, the marginals of derived quantities of interest such as the dynamic pressure and the heat rate are computed, exploiting the already derived marginal distributions of the propagated variables. Finally, we use the derived marginals to assess the compliance of the mission to relevant constraints.

% --------------------------------------------------------------
% TEST CASE - EARTH 3 STATE
% --------------------------------------------------------------
\subsection{Three-state steep Earth re-entry}  \label{subsec:test_strategic}
The following section features a so-called \emph{strategic} re-entry \citep{Putnam2015}, which represents a vehicle with a high ballistic coefficient on a steep, high-energy trajectory. The aim of these types of re-entry is to more precisely pinpoint the impact location on Earth. The test case is performed utilizing the \emph{three-state} dynamics of \cref{eq:system3state}. For the modeling of the atmosphere, a simple exponential model is adopted (\cref{eq:atm}), while for the gravitational acceleration an inverse square model is considered (\cref{eq:grav}). The expressions for the two models are as follows:

\begin{align}
\rho(r) &= \rho_0  \exp{ \frac{H_2 - (r - R_p)}{H_1} }  \label{eq:atm} \\
g(r) &= \frac{\mu_p}{r^2} \label{eq:grav}
\end{align}

where $\rho_0$ is a reference atmospheric density, $H_1$ and $H_2$ are constants related to the atmosphere of the planet, and $\mu_p$ is the gravitational parameter of the planet. The initial conditions of the re-entry together with their uncertainties and the parameters used for the models are summarized in \cref{tab:strategic_state0}. The uncertainties have been modeled with a Gaussian distribution so that the values of the initial conditions of \cref{tab:strategic_state0}, $h_0$, $v_0$, $\gamma_0$, and $\beta_0$ represent the mean of a multivariate Gaussian.

\begin{table}[hbt!]
	\caption{\label{tab:strategic_state0} Initial conditions and relevant parameter for the three-state strategic re-entry on Earth.}
	\centering
	\begin{tabular}{lcccc}
		\hline
		State                     & Symbol      & Unit                               & $\mu$ & $\sigma$ \\
		\hline
		Initial velocity          & $v_0$       & \si{\kilo\meter / \second}         & 7.2   & 5 \\
		Initial flight path angle & $\gamma_0$  & \si{\degree}                       & -30.0 & 0.1 \\
		Initial altitude          & $h_0$       & \si{\kilo\meter}                   & 125   & 2 \\
		Ballistic coefficient     & $\beta_0$   & \si{\kilo\gram / \square\meter} 	 & 10000 & 500 \\
		\hline
		\hline
		Parameter                      & Symbol    & Unit                                 & Value                    & \\
		\hline
		Earth equatorial radius        & $R_p$     & \si{\kilo\meter}                     & 6378.1                   & \\
		Earth gravitational parameter  & $\mu_p$   & \si{\cubic\meter / \second\squared}  & 3.986$\times 10^{14}$ & \\
		Reference atmospheric density  & $\rho_0$  & \si{\kilo\gram / \cubic\meter}       & 1.215                    & \\
		Atmospheric scale height       & $H_1$     & \si{\kilo\meter}                     & 8.3                      & \\
		Reference altitude             & $H_2$     & \si{\kilo\meter}                     & 0                        & \\
		Lift-to-drag ratio             & $C_L/C_D$ &                                      & 0                        & \\
		\hline
	\end{tabular}
\end{table}

\begin{figure}[hbt!]
	\centering
	\includegraphics[width=0.6\textwidth]{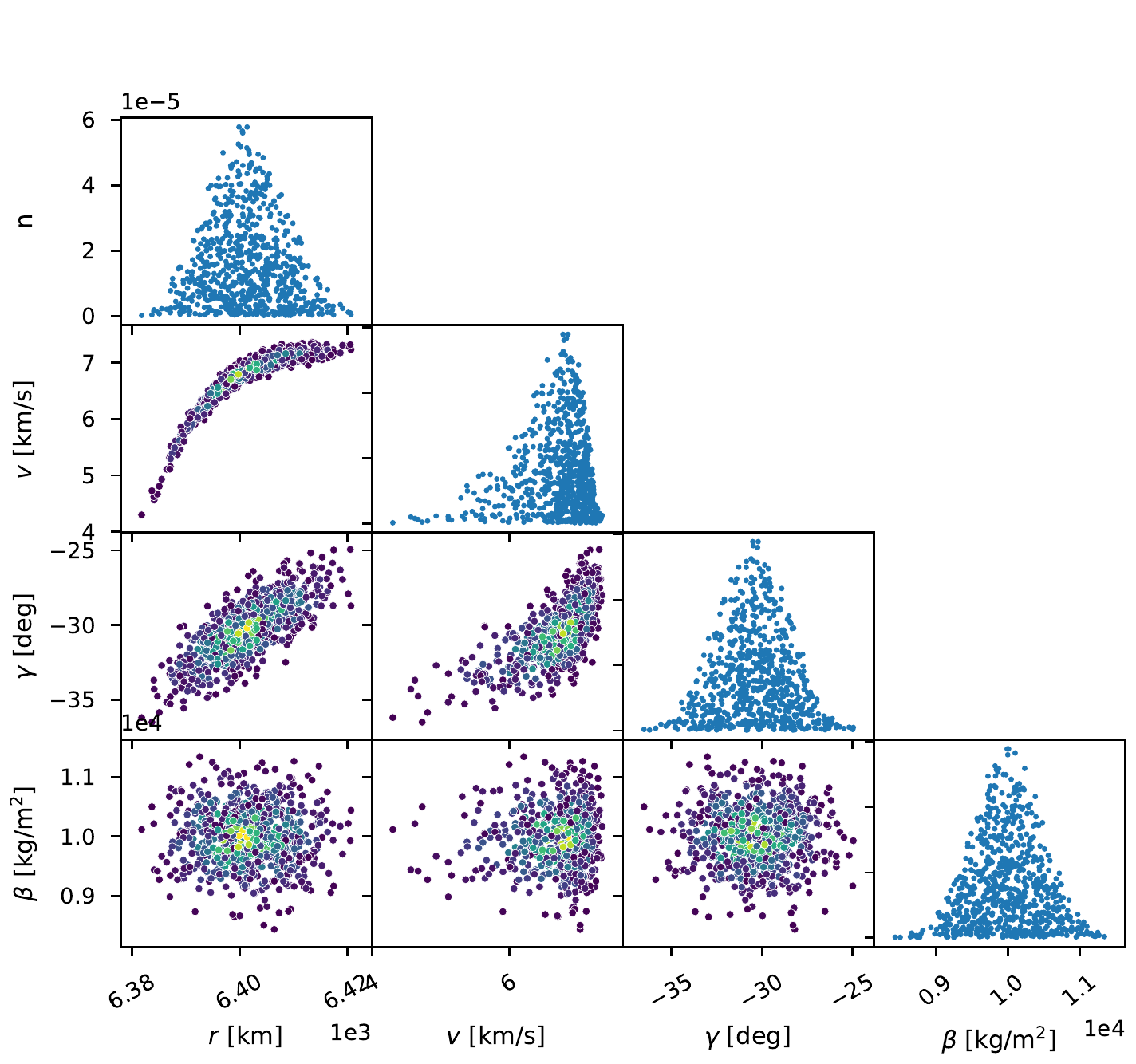}
	\caption{Pairplot of the  probability density propagation for the strategic re-entry at two different snapshots at $\boldsymbol{t = 30}$ s.}
	\label{fig:strategic_scatter}
\end{figure}

For the case in exam, the uncertainties affect the initial state of the satellite at the re-entry interface ($r$, $v$, $\gamma$) and the ballistic coefficient ($\beta$). No uncertainty in the atmosphere is considered for this case so that the $\xi$ coefficient is neglected. We decided to consider the ballistic coefficient as it can influence the shape and evolution of the re-entry trajectory and the prediction of the landing location. In addition, the exact quantification of the ballistic coefficient at re-entry can be difficult as it depends on the mass, cross-section and drag coefficient of the spacecraft. The mass of the spacecraft at disposal may not be completely known, as the exact amount of residual propellant can be difficult to assess. The cross-sectional area of the satellite may be also uncertain as well as even the exact value of the drag coefficient. Following the procedure outlined in \cref{sec:propagation}, once the initial uncertainty distribution has been defined, this is sampled and each sample is propagated using \cref{eq:system3state}. In this case, we draw 750 samples. The result of the propagation using the continuity equation for selected snapshots (time instants, counted in seconds, after the starting epoch of the simulation) are summarized in \cref{fig:strategic_scatter}.

The pairplot of \cref{fig:strategic_scatter} shows the projection of the sampled points for the snapshot at time $t=30$ s. Along the main diagonal, the value of the density ($n$) associated to each sample is presented for each variable: the diagonal plot in the first column shows the distribution of the samples in $r$ and the associated density value, the one in the second column refers to the velocity $v$, the one in the third column refers to the flight-path angle $\gamma$, and the one in the last column gives the density distribution as a function of the ballistic coefficient $\beta$. The off diagonal two-dimensional plots show the relation between different couples of variables. The color map of the two-dimensional plots is associated to the magnitude of the density of each point. It is possible to observe how the state space, which started from a Gaussian structure, suffered a substantial deformation as the time passed. The deformation represents one of the challenges in reconstructing the probability density for the entire state-space volume.

\subsubsection{Comparison}  \label{subsubsec:earth3state_results}
\begin{figure}[hbt!]
	\centering
	\begin{subfigure}[b]{0.32\textwidth}
		\includegraphics[height=4.6cm]{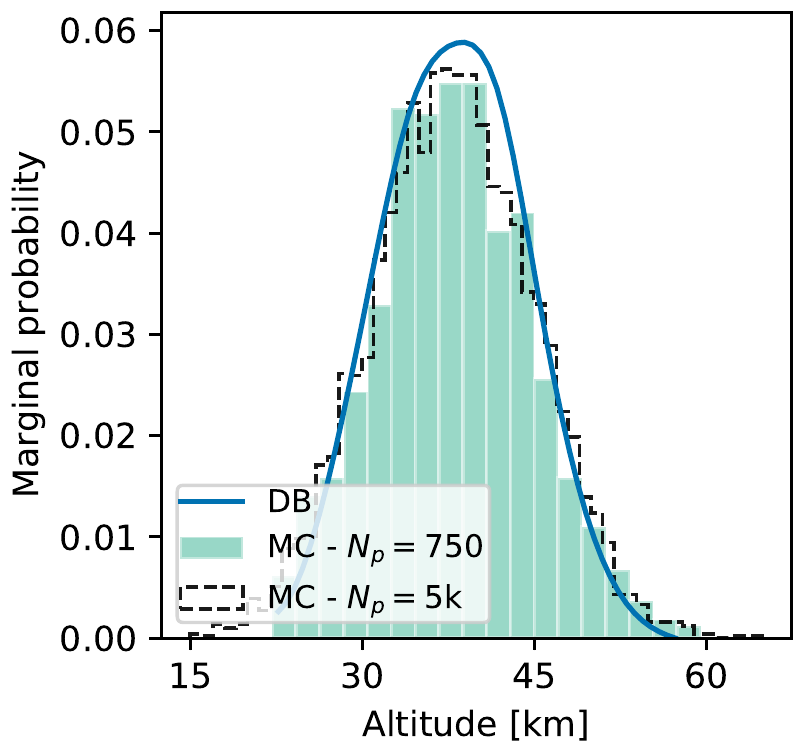}
		\caption{ }
		\label{fig:h_24_750}
	\end{subfigure}
	~
	\begin{subfigure}[b]{0.32\textwidth}
		\includegraphics[height=4.6cm]{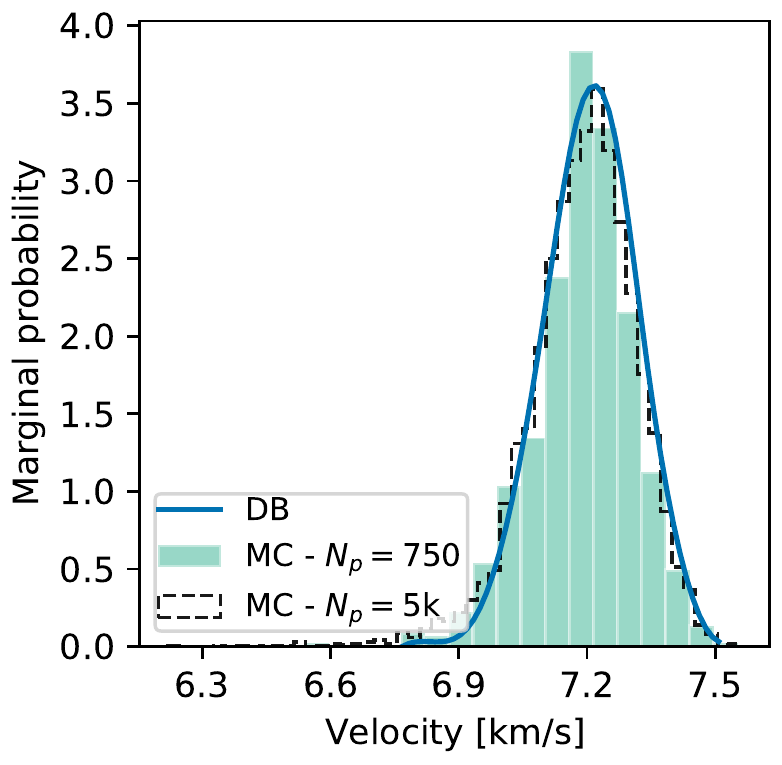}
		\caption{ }
		\label{fig:v_24_750}
	\end{subfigure}
	~
	\begin{subfigure}[b]{0.32\textwidth}
		\includegraphics[height=4.6cm]{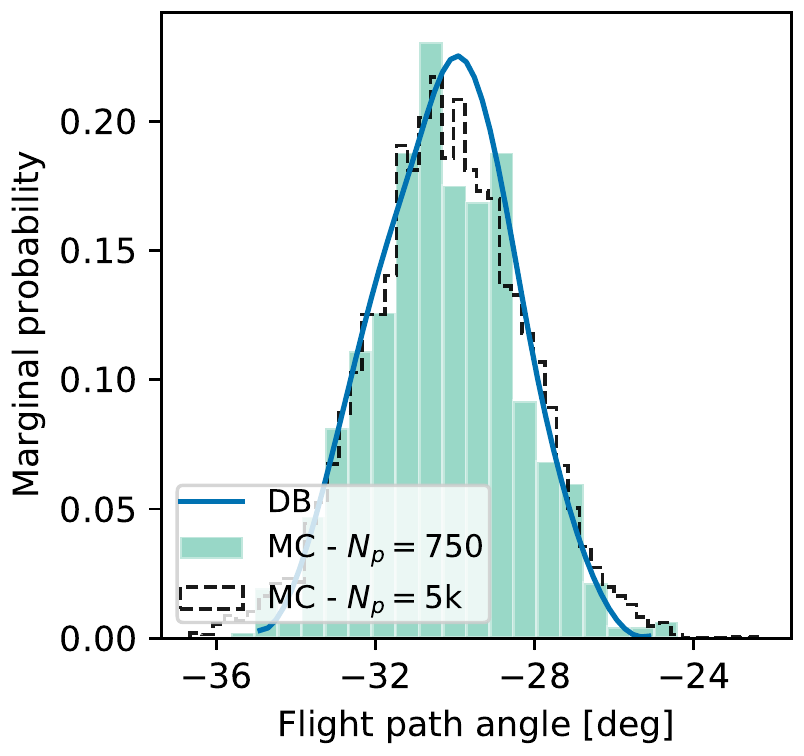}
		\caption{ }
		\label{fig:fpa_24_750}
	\end{subfigure}
	\caption{DB vs. MC one-dimensional marginals comparison at $\boldsymbol{t = 24}$ s. DB: marginals with 750 samples. MC: marginals with 750 and 5000 samples.}
	\label{fig:m1d_earth3state}
\end{figure}

\cref{fig:h_24_750,fig:v_24_750,fig:fpa_24_750} show the one-dimensional marginals obtained with the density-based approach (DB) and the Monte Carlo sampling (MC) for the snapshot at time $t=24$ s. In this comparison we have used 750 samples for the DB method and 750 (shaded histogram) and 5000 (dashed histogram) samples for the MC method. 

\begin{figure}[hbt!]
	\centering
	\begin{subfigure}[b]{0.40\textwidth}
		\includegraphics[height=4.1cm]{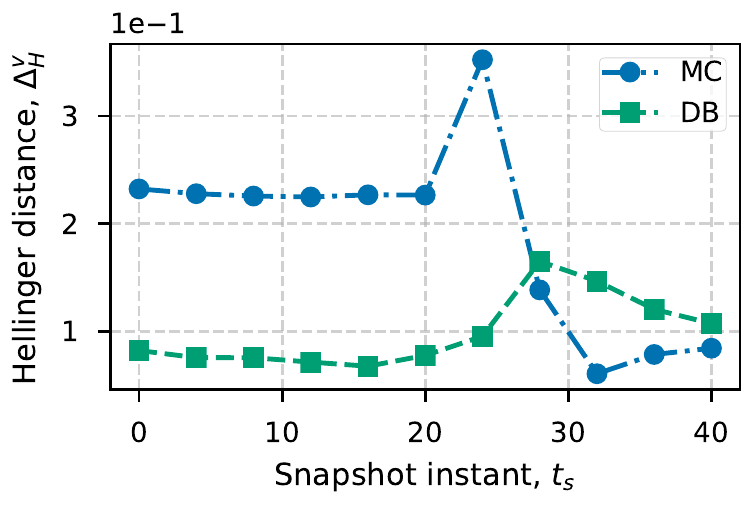}
		\caption{ }
		\label{fig:deltaH_v}
	\end{subfigure}
	\quad\quad
	\begin{subfigure}[b]{0.40\textwidth}
		\includegraphics[height=4.1cm]{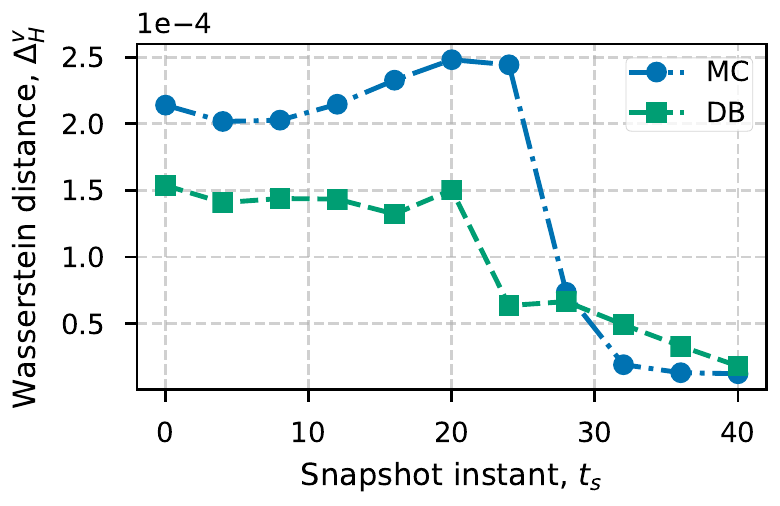}
		\caption{ }
		\label{fig:deltaW_v}
	\end{subfigure}
	\caption{Variation of the Hellinger distance (a) and Wasserstein distance (b) in time for the velocity. Both distances compare the 750 samples marginals of the DB and MC methods to the 5000 MC samples.}
	\label{fig:comparison_3state}
\end{figure}

To better compare the obtained marginal, we measure the difference between the distributions obtained with the DB and MC methods. To do so, we introduce two metrics commonly used in statistics to compare probability distributions that are the Hellinger distance ($\Delta_H$) and the first Wasserstein distance ($\Delta_W$). The Hellinger distance and the first Wasserstein distance between two PDFs $\mathcal{P}(x)$ and $\mathcal{Q}(x)$ are defined as follows:

\begin{align}
\Delta_{H}(\mathcal{P}, \mathcal{Q}) &= \sqrt{ \frac{1}{2} \int_{-\infty}^{+\infty} (\sqrt{\mathcal{P}} - \sqrt{\mathcal{Q}})^2 dx }  \\
\Delta_{W}(\mathcal{P}, \mathcal{Q}) &= \inf_{\pi \in \Gamma(\mathcal{P}, \mathcal{Q})} \int \|x - y\| d\pi(x, y)
\end{align}

where $\Gamma(\mathcal{P}, \mathcal{Q})$ is the collection of measures with marginals $\mathcal{P}$ and $\mathcal{Q}$. The Hellinger distance is a form of divergence measuring the difference between two distributions and always lies between 0 and 1. The Wesserstein distance can be interpreted as the amount of \emph{work} needed to transform one distribution into the other. \cref{fig:deltaH_v,fig:deltaW_v} show the value of the Hellinger and Wesserstein distance respectively as a function of time (i.e. different snapshots fitting) for the variable $v$, comparing the DB and MC distributions with 750 to a reference distribution that is the MC marginal obtained with 5000 samples.

It is possible to observe that the DB method shows a better performance in both metrics from the initial states until about 28 seconds into the re-entry. From that point on, the MC method performs slightly better. This is probably due to the characteristics of the simulations; in fact, after 30 seconds some of the samples reach the ground and are thus no longer used in the interpolation and therefore a loss of performance can be expected (as shown by the Hellinger distance). On the other hand, it is surprising that the MC method considerably improve in both metrics when less points are available. The comparison after this instant might be less accurate given that also the reference distribution obtained with 5000 samples suffers from a proportional reduction in the number of points available. Nonetheless, we show in \cref{tab:comparison3state}, the comparison between the average value and the standard deviation over all the snapshots of the Hellinger and Wesserstein distances for $r$, $v$, and $\gamma$ for the DB and MC methods.

\begin{table}[hbt!]
	\caption{\label{tab:comparison3state} Comparison of mean and standard deviation of the Hellinger and Wasserstein distances for the considered states.}
	\centering
	\begin{tabular}{l|cc|cc|cc}
		& \multicolumn{2}{c}{$r$} & \multicolumn{2}{c}{$v$} & \multicolumn{2}{c}{$\gamma$} \\
		\hline
		& $\mu_{\Delta_H}$ & $\sigma_{\Delta_H}$ & $\mu_{\Delta_H}$ & $\sigma_{\Delta_H}$ & $\mu_{\Delta_H}$ & $\sigma_{\Delta_H}$ \\
		\hline
		DB     & 0.1354 & 0.0327 & 0.0986 & 0.0313 & 0.0967 & 0.0064 \\
		MC     & 0.1679 & 0.0623 & 0.1889 & 0.0840 & 0.1195 & 0.0095 \\
		\hline
		\hline
		& $\mu_{\Delta_W}$ & $\sigma_{\Delta_W}$ & $\mu_{\Delta_W}$ & $\sigma_{\Delta_W}$ & $\mu_{\Delta_W}$ & $\sigma_{\Delta_W}$ \\
		\hline
		DB     & \num{3.12e-6} & \num{8.20e-7} & \num{9.95e-5} & \num{5.06e-5} & \num{4.80e-1} & \num{1.02e-1} \\
		MC     & \num{4.93e-6} & \num{1.32e-6} & \num{1.52e-6} & \num{9.52e-5} & \num{6.94e-1} & \num{1.64e-1} \\
		\hline
	\end{tabular}
\end{table}

It can be observed that both metrics are, on average, lower for the DB method as opposed to the MC method for the same number of samples, and the standard deviation is also smaller in all instances, showing that the accuracy of the proposed method is competitive with MC using an equivalent sample size. A comparison of the computational effort for the strategic test case is presented in \cref{tab:time_strategic_case} for the DB case with 750 samples and the MC case with 750 and 5000 samples respectively. The computational times have been subdivided into propagation time that is the time needed to propagate the trajectories and the marginalisation time that is the time needed to reconstruct the marginal distribution from the scattered samples. The simulations were ran on a Windows laptop equipped with an Intel Core i7-1065G7 @ 1.3 GHz processor and 16 GB of RAM.

\begin{table}[hbt!]
	\caption{\label{tab:time_strategic_case} Runtime comparison between the DB and MC methods for the strategic test case.}
	\centering
	\begin{tabular}{lcc|c}
		\hline
		Case 			& Propagation time (s) 	& Marginalisation time (s) 	& Total time (s) 	\\
		\hline
		DB - Np = 750 	& 8.2 					& 1.2 						& 9.4				\\
		MC - Np = 750  	& 8.2 					& 0.01						& 8.21				\\
		MC - Np = 5000  & 74.0 					& 0.06 						& 74.06				\\
		\hline
	\end{tabular}
\end{table}

\begin{figure}[hbt!]
	\centering
	\begin{subfigure}[b]{0.32\textwidth}
		\includegraphics[width=\textwidth]{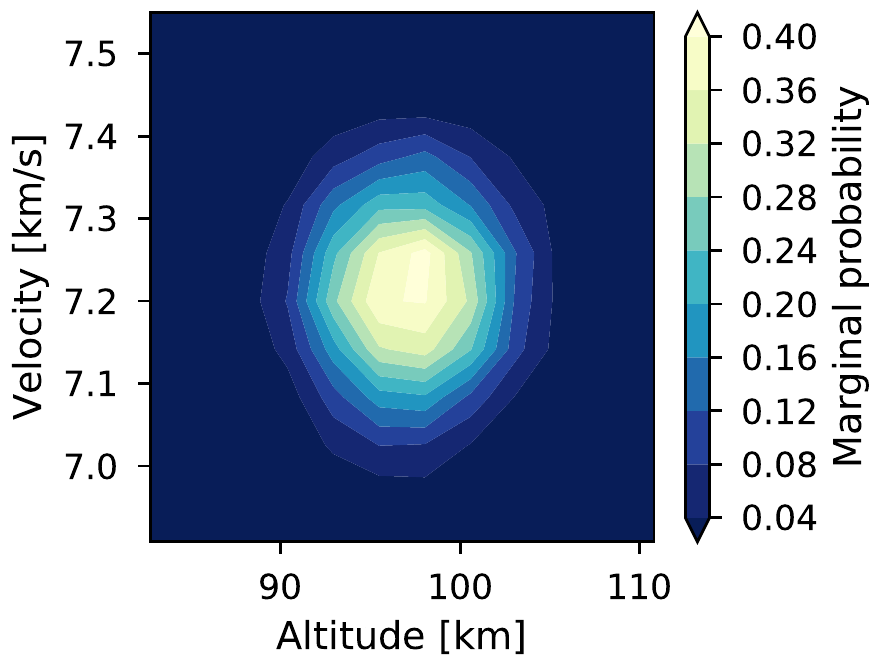}
		\caption{DB method with 750 samples.}
		\label{fig:hv_DB_8_750}
	\end{subfigure}
	~
	\begin{subfigure}[b]{0.32\textwidth}
		\includegraphics[width=\textwidth]{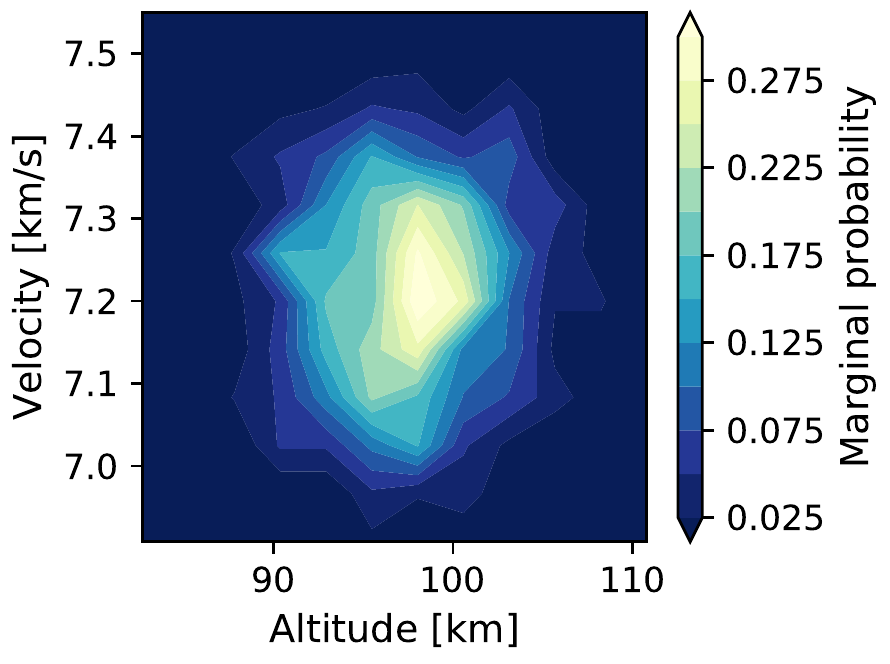}
		\caption{MC method with 750 samples.}
		\label{fig:hv_MC_8_750}
	\end{subfigure}
	~
	\begin{subfigure}[b]{0.32\textwidth}
		\includegraphics[width=\textwidth]{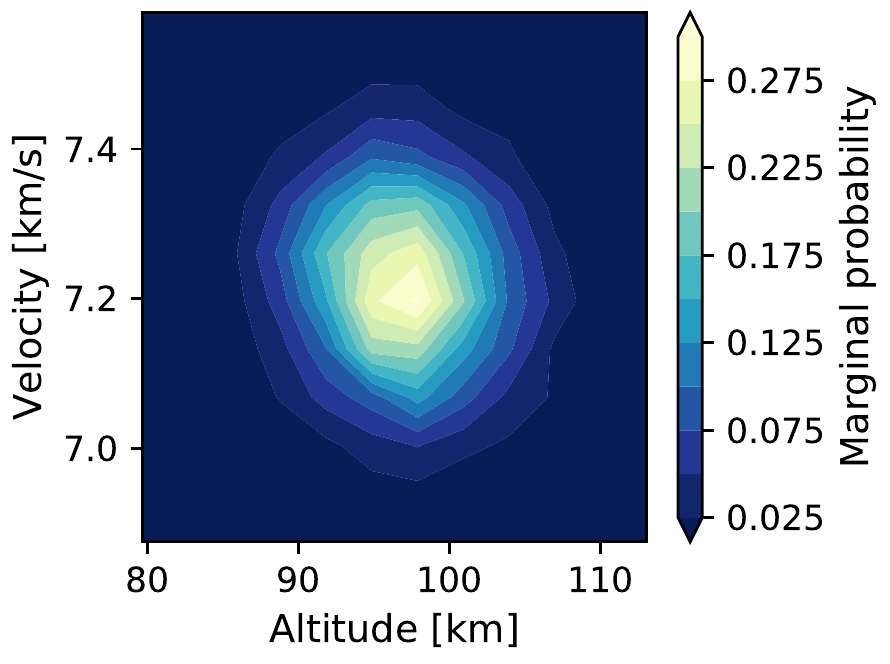}
		\caption{MC method with 5000 samples.}
		\label{fig:hv_MC_8_5k}
	\end{subfigure} \\
	\begin{subfigure}[b]{0.32\textwidth}
		\includegraphics[width=\textwidth]{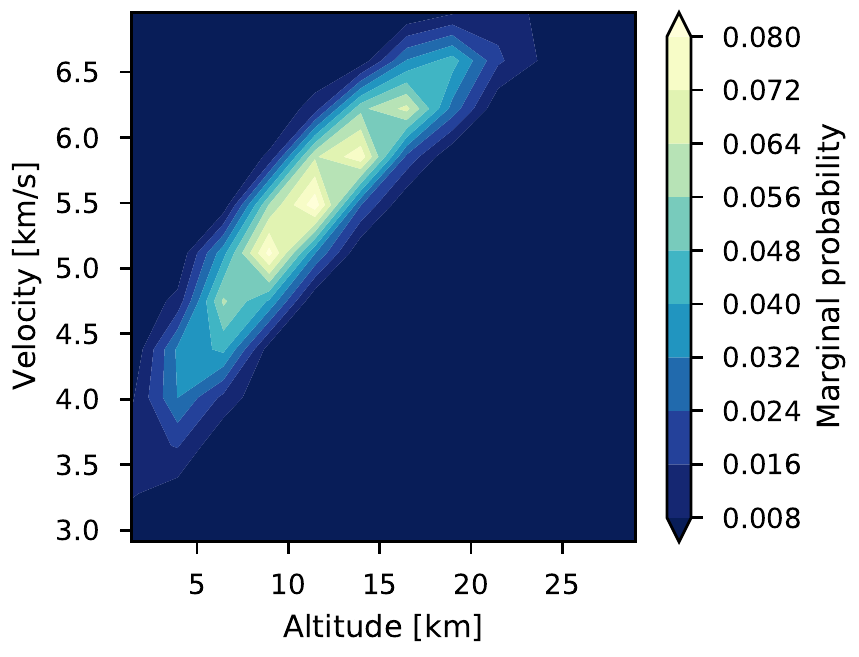}
		\caption{DB method with 750 samples.}
		\label{fig:hv_DB_32_750}
	\end{subfigure}
	~
	\begin{subfigure}[b]{0.32\textwidth}
		\includegraphics[width=\textwidth]{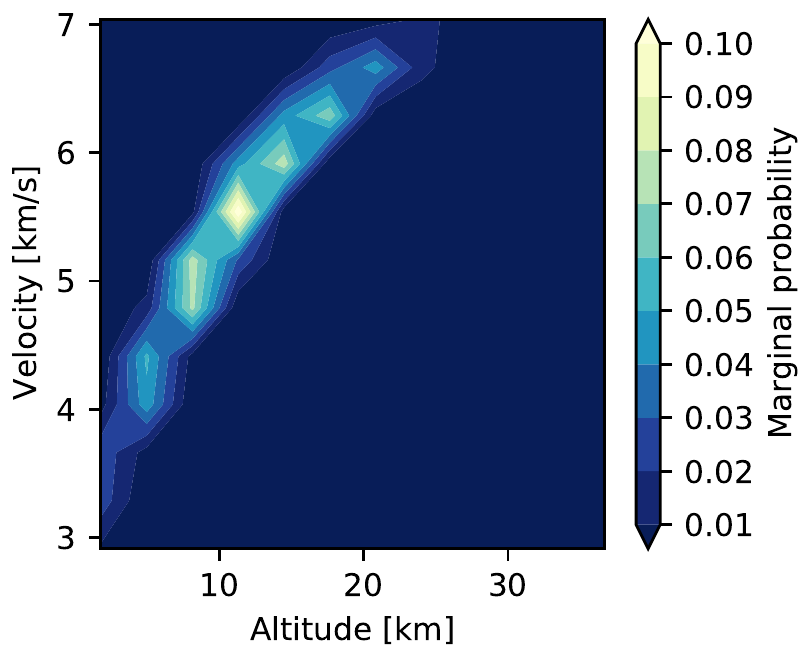}
		\caption{MC method with 750 samples.}
		\label{fig:hv_MC_32_750}
	\end{subfigure}
	~
	\begin{subfigure}[b]{0.32\textwidth}
		\includegraphics[width=\textwidth]{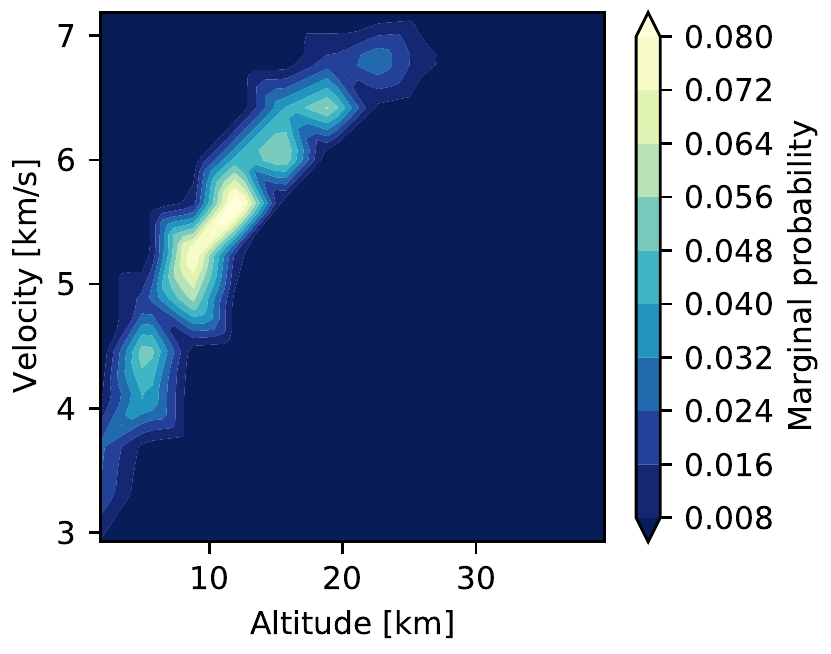}
		\caption{MC method with 5000 samples.}
		\label{fig:hv_MC_32_5k}
	\end{subfigure}
	
	\caption{DB vs. MC two-dimensional marginals in $\boldsymbol{h}$-$\boldsymbol{v}$. Top row: snapshot time $\boldsymbol{t = 8}$ s. Bottom row: snapshot time $\boldsymbol{t = 32}$ s.}  
	\label{fig:strategic_2d_marginals}
\end{figure}

\cref{fig:strategic_2d_marginals} shows a visual comparison between the two-dimensional marginals in altitude and velocity for two distinct snapshots ($t=8$ s and $t=32$ s), obtained with the DB approach using 750 samples (\cref{fig:hv_DB_8_750,fig:hv_DB_32_750}), and the MC approach using 750 and 5000 samples respectively (\cref{fig:hv_MC_8_750,fig:hv_MC_8_5k,fig:hv_MC_32_750,fig:hv_MC_32_5k}). 

A feature that can be observed in the obtained marginals (\cref{fig:m1d_earth3state,fig:strategic_2d_marginals}) is that the DB approach tends to produce \emph{more localized} distributions. On one hand, the DB methodology uses the actual value of the probability density to obtain the marginals, while Monte Carlo estimates them and may under-predict or over-predict the distributions. However, the reconstructed marginals with the DB method are only as good as the fitting procedure used and the integration performed. The marginalization procedure proposed in \cref{subsec:marginals} together with the interpolation presented in \cref{subsec:dual-taylor} introduces some simplifications and is limited to the envelope of the state space defined by the sampled points, which are then used to construct the $\alpha$-shape. This feature must be monitored as it may arise from a lower accuracy that can be obtained with the DB method at the boundaries of the distribution where a lower density of samples is available. An issue that might be mitigated by a more spatially-efficient sampling technique. 

\begin{figure}[hbt!]
	\centering
	\includegraphics[width=0.45\textwidth]{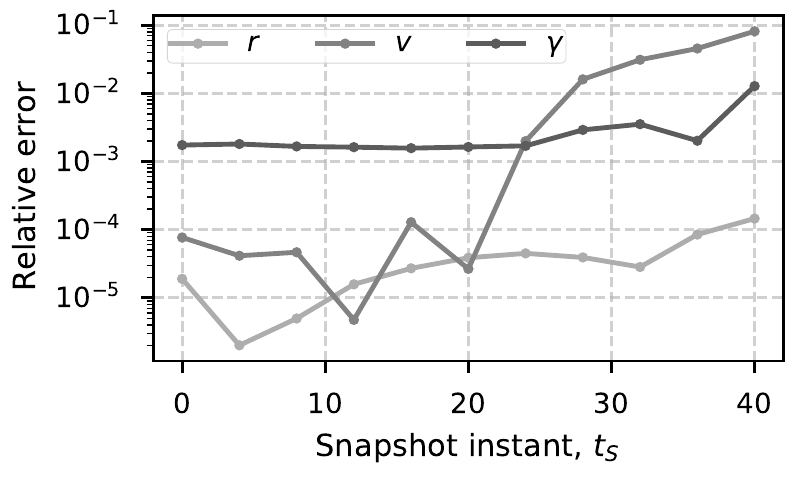}
	\caption{Evolution of the relative difference between the expected value predicted by the DB method (750 samples) and MC method (5000 samples) for $\boldsymbol{r}$, $\boldsymbol{v}$, and $\boldsymbol{\gamma}$.}
	\label{fig:strategic_relerr}
\end{figure}

\cref{fig:strategic_relerr} shows the relative difference between the mean value predicted by the DB method with 750 samples and the MC method with 5000 samples at each snapshot for the three variables of interest ($r$, $v$, $\gamma$). It can be observed from \cref{fig:strategic_relerr} that the difference between the mean value as predicted by the MC and the DB methods remains below 1\% for both the radius and flight-path angle, while it increases up to 8\% for the velocity in the last part of the trajectory. This behavior confirms that in the last part of the trajectory, when fewer samples are available (in the last instant only 110 samples out of 750 are available), the performance deteriorates as expected.

\begin{figure}[hbt!]
	\centering
	\begin{subfigure}[b]{0.32\textwidth}
		\includegraphics[height=5.1cm]{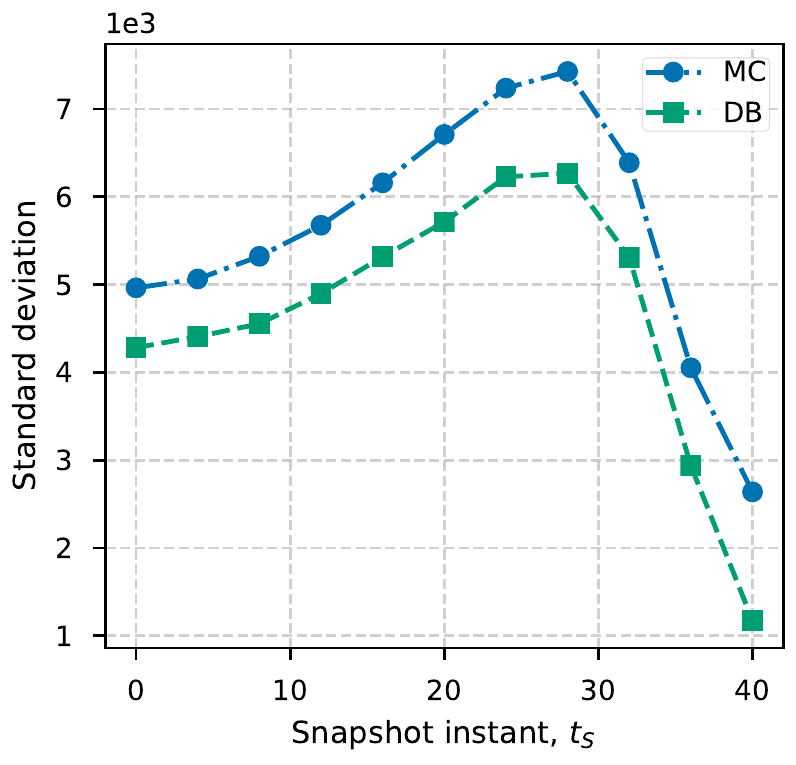}
		\caption{ }
		\label{fig:strategic_r_std}
	\end{subfigure}
	~
	\begin{subfigure}[b]{0.32\textwidth}
		\includegraphics[height=5.1cm]{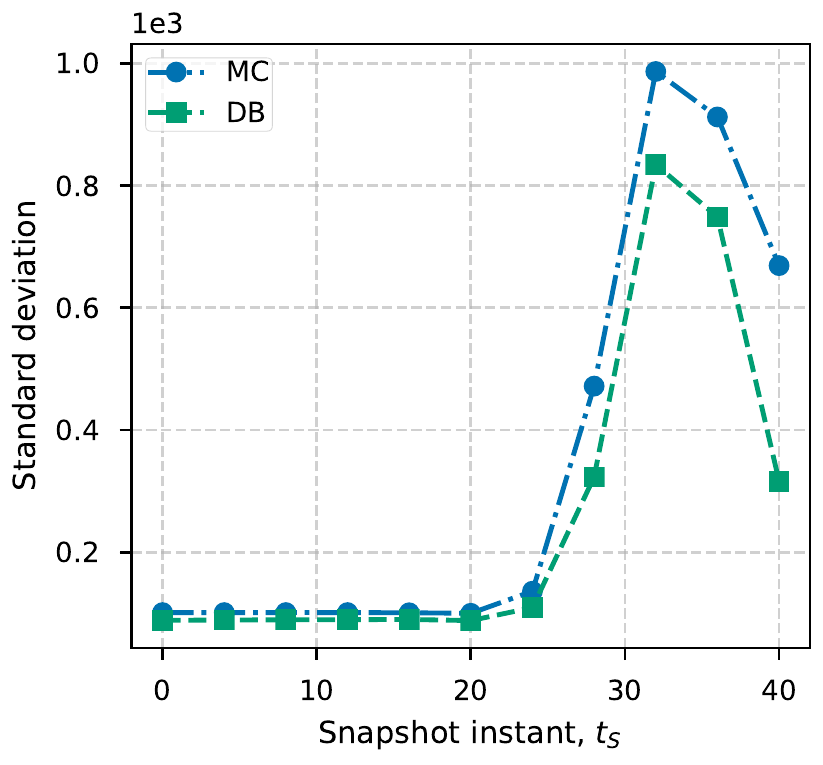}
		\caption{ }
		\label{fig:strategic_v_std}
	\end{subfigure}
	~
	\begin{subfigure}[b]{0.32\textwidth}
		\includegraphics[height=5.1cm]{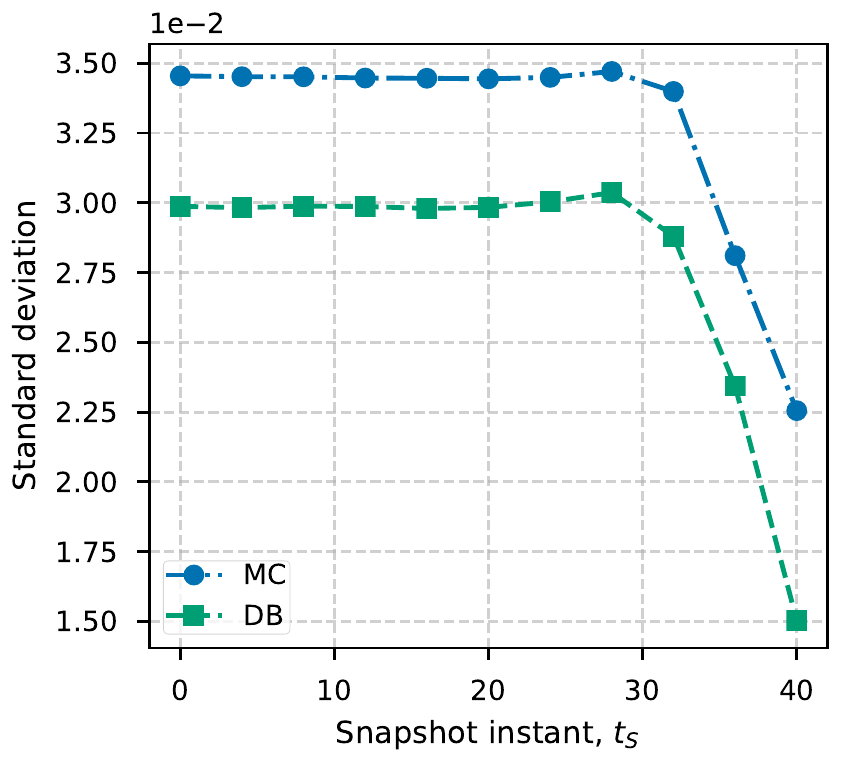}
		\caption{ }
		\label{fig:strategic_fpa_std}
	\end{subfigure}
	\caption{Evolution of the standard deviation predicted by the DB method (750 samples) and MC method (5000 samples) for $\boldsymbol{r}$ (a), $\boldsymbol{v}$ (b), and $\boldsymbol{\gamma}$ (c).}
	\label{fig:strategic_std_comparison}
\end{figure}

\cref{fig:strategic_std_comparison} shows instead the difference between the standard deviation predicted by the DB method with 750 samples and the MC method with 5000 samples for each snapshot. The results shows that the standard deviation predicted by the DB method follows the one estimated with the MC method. It also confirms what could be observed in \cref{fig:m1d_earth3state} and \cref{fig:strategic_2d_marginals} with the DB method tendency to underpredict the tails of the distributions.

\subsubsection{Mechanical and thermal loads compliance}  \label{subsubsec:strategic_compliance}
During re-entry assessments it is not only of interest to predict the uncertainties in the position and velocity of the spacecraft, but also to understand how they transfer to other relevant quantities such as the dynamic pressure and the heat rate. We compute these quantities using the following expressions:

\begin{align}
\bar{q} &= \frac{1}{2} \; \rho(h) \; v^2 \label{eq:pdyn} \\
\dot{Q} &= \Bar{F}_{\rm q} \; K \; \sqrt{\frac{0.3048}{r_n}} \sqrt{\frac{\rho(h)}{\rho_{\rm SL}}} \; \bigg( \frac{v}{7924.8} \bigg)^{3.15},  \label{eq:qdot}
\end{align}

where $K=1.99876 \times 10^8$ \si{\watt / \square\meter} is a constant, $r_n$ is the curvature radius at stagnation point in meters, $\rho_{\rm SL}$ is the air density at sea-level in \si{\kilo\gram / \cubic\meter}, and $\Bar{F}_{\rm q}$ is an averaging factor that depends on the shape of the object, its motion, and the re-entry regime. The expression for the heat rate is a simplified version of the Detra-Kemp-Riddell correlation (DKR) \citep{kemp1957heat}, where we have omitted the term related to the stagnation point enthalpy. The DKR correlation is commonly used to predict the heat rate for hypersonic entries in Earth atmosphere \citep{Beck2015,TLC18AESCTE,Gelhaus2014}. It is applicable to continuum flows so that it is less accurate at higher altitudes; however, for this demonstrative example it has been used to compute the heat rate for the entire trajectory. 

If we select the characteristics of the spacecraft, i.e. the curvature radius at the stagnation point and the averaging factor, both \cref{eq:pdyn,eq:qdot} are only a function of the altitude and the velocity. We can thus exploit the previously computed marginal distributions in $h$ and $v$ to directly obtain information on the dynamic pressure and the heat rate during re-entry. In fact, we can apply a transformation and scale the density accordingly. For example, in the case of the heat rate we have:

\begin{equation}  \label{eq:transform}
\begin{cases}
\varphi(h, v) = v \\
\psi(h, v) = \Bar{F}_{\rm q} \; K \; \sqrt{\frac{0.3040}{r_n}} \sqrt{\frac{\rho(h)}{\rho_{\rm SL}}} \; \big( \frac{v}{7924.8} \big)^{3.15}.
\end{cases}
\end{equation}

When we apply this transformation, we need to scale the density accordingly, which can be accomplished by dividing the density in the original variables by the determinant of the Jacobian of the transformation as follows:

\begin{equation}  \label{eq:dens_scale}
n(v, \dot{Q}) = n(h, v) \; 
\begin{vmatrix}
\pdv{\varphi}{h} & \pdv{\varphi}{v} \\
\pdv{\psi}{h} & \pdv{\psi}{v}
\end{vmatrix}^{-1}.
\end{equation}

\begin{figure}[hbt!]
	\centering
	\begin{subfigure}[b]{0.48\textwidth}
		\centering
		\includegraphics[height=5.6cm]{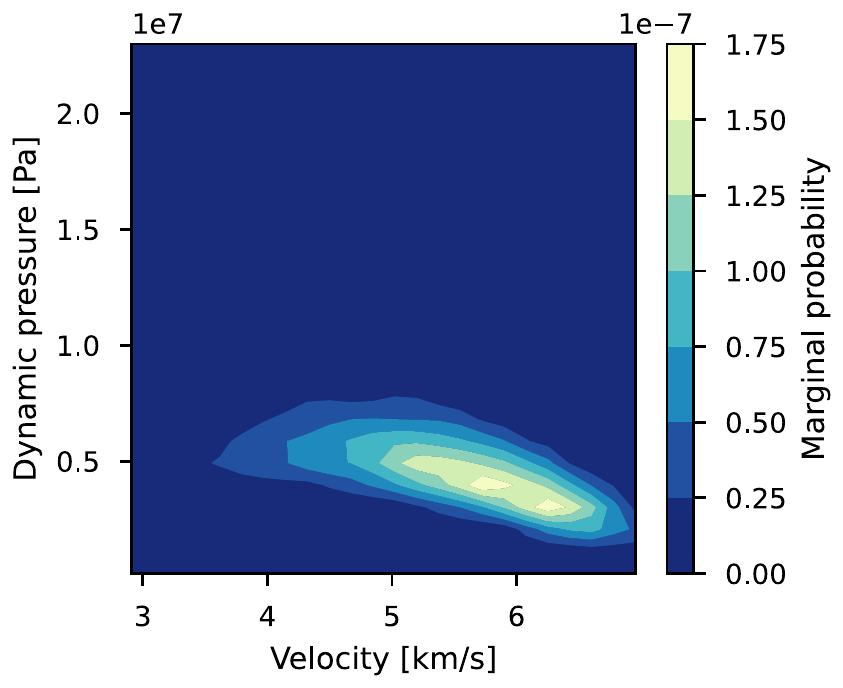}
		\caption{ }
		\label{fig:pdyn_1}
	\end{subfigure}
	~
	\begin{subfigure}[b]{0.48\textwidth}
		\centering
		\includegraphics[height=5.6cm]{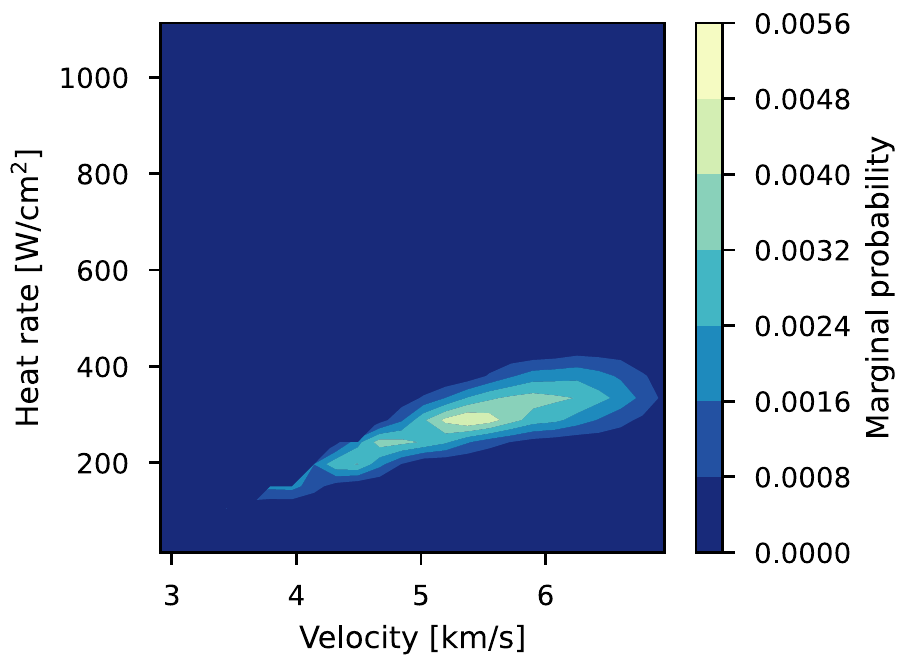}
		\caption{ }
		\label{fig:qdot_2}
	\end{subfigure}
	\caption{Two-dimensional DB marginals of velocity vs. dynamic pressure (a) and velocity vs heat rate (b) at time $\boldsymbol{t = 32}$ s.}
\end{figure}

With an equivalent procedure, we can also obtain the marginals for the dynamic pressure. \cref{fig:pdyn_1,fig:qdot_2} show examples for both the dynamic pressure and the heat rate for the test case in exam. The heat rate has been obtained considering a spherical shape with a one meter radius. The corresponding averaging factor is $\Bar{F}_{\rm q} = 0.234$ \citep{TLC18Sensitivity}.

\begin{figure}[hbt!]
	\centering
	\begin{subfigure}[b]{0.45\textwidth}
		\centering
		\includegraphics[height=6cm]{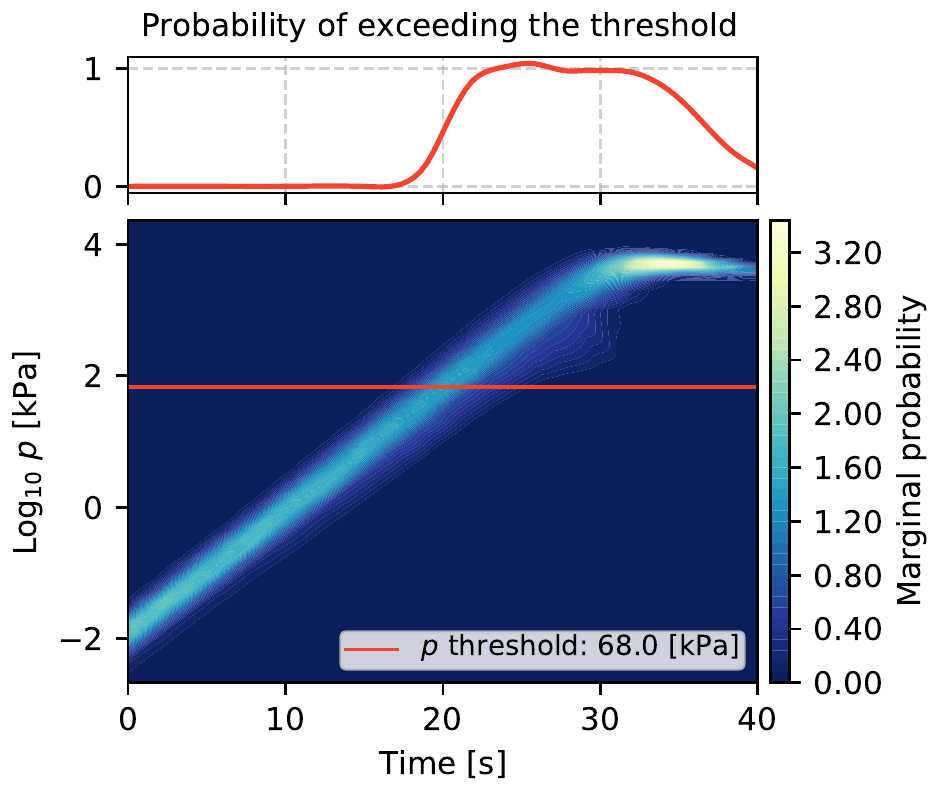}
		\caption{ }
		\label{fig:p_time}
	\end{subfigure}
	~
	\begin{subfigure}[b]{0.45\textwidth}
		\centering
		\includegraphics[height=6cm]{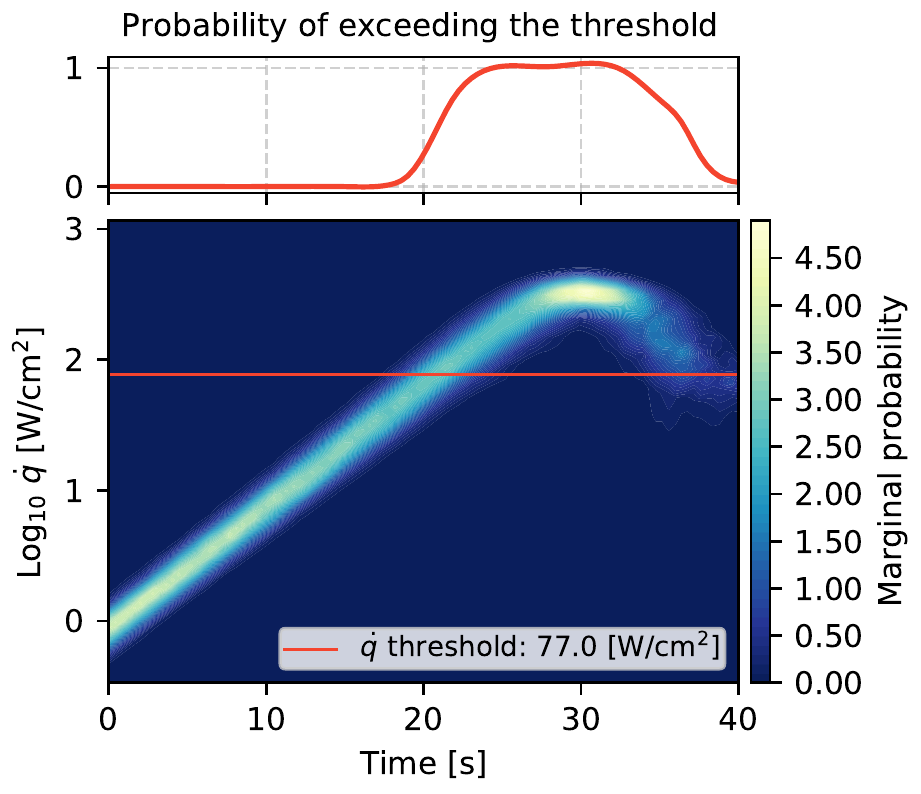}
		\caption{ }
		\label{fig:qdot_time}
	\end{subfigure}
	\caption{Dynamic pressure (a) and heat rate (b) marginal distributions evolution in time. In red, specific thresholds. On top, the probability in time of crossing the threshold.}
\end{figure}

These two-dimensional marginals can then be integrated along $v$ to obtain the one-dimensional marginals for the dynamic pressure and heat rate. We can combine them together and obtain the evolution in time of the marginal probability for both the dynamic pressure and the heat rate. \cref{fig:p_time,fig:qdot_time} show this time evolution. Each plot also features a threshold (in red). These values can be set by the user and defined to check different scenarios. For example, the limits in dynamic pressure and heat rate that a re-entry capsule can withstand. For this test case, a threshold of 68 \si{\kilo\pascal} and 77 \si{\watt / \centi\meter\squared} \citep{lu2014entry} for the dynamic pressure and heat rate, respectively. The top part of the plot, represents the probability at each time step of crossing the limit. Initially, no part of the dynamic pressure and heat rate distributions crosses the thresholds. As time passes, the distributions shift towards higher dynamic pressures and heat rates so that parts of the distribution cross the threshold, increasing the probability.  In the final states, the probability density starts to reduce again. This happens because some of the initial samples reach the ground before others and when they do, their contribution to the total probability weight is removed.  

% --------------------------------------------------------------
% TEST CASE - EARTH 6 STATE
% --------------------------------------------------------------
\subsection{Six-state steep Earth re-entry}  \label{subsec:test_6stateEarth}
The example presented in the following section is an extension of the test case in \cref{subsec:test_strategic} in which a six-state propagation is used instead (\cref{eq:system6state}) and the gravity and atmosphere models have an increased complexity. The gravitational model used takes into account the J2 effect \citep{Tewari2007} and the atmospheric model is the 1976 US Standard Atmosphere \citep{atmosphere1976}. In order to use the US Standard Atmosphere model into the continuity equation, a spline interpolation of the density has been performed using the \emph{UnivariateSpline} class of the \emph{scipy} Python package. The \emph{UnivariateSpline} class has the advantage of providing the derivative information that can be used directly into the equations of motion when needed.
As previously mentioned, the uncertainties in the atmosphere can be included in the continuum approach. The procedure adopted here is to introduce a correction coefficient to scale the value of the density provided by the selected model as follows:

\begin{equation}  \label{eq:atmCorrection}
\rho(r, \xi) = \xi \; \hat{\rho}_{\rm std76}(r),
\end{equation}

where $\xi$ is the atmospheric correction coefficient and $\hat{\rho}_{\rm std76}(r)$ is the spline representation of the atmospheric density as provided by the 1976 US Standard Atmosphere model. More complex models, with several uncertain coefficients can be included, as long as they can be expressed with differentiable functions and that they are added to the augmented state space during the continuum propagation. For the case in exam, the uncertainties have been considered in all the initial states ($\lambda$, $\varphi$, $v$, $\gamma$, $\chi$) and in the atmospheric correction coefficient ($\xi$), while no uncertainty in the ballistic coefficient has been included. \cref{tab:earth6state0} summarizes the mean value and standard deviation for the considered uncertain states and the value of the remaining parameters necessary for the re-entry simulation.

\begin{table}[hbt!]
	\caption{\label{tab:earth6state0} Mean and standard deviation for the initial conditions of the augmented state space and values of the relevant parameters.}
	\centering
	\begin{tabular}{lcccc}
		\hline
		State                     & Symbol      & Unit                               & $\mu$ & $\sigma$ \\
		\hline
		Initial longitude         & $\lambda_0$ & \si{\degree}                       & 0     & 0.2 \\
		Initial latitude          & $\varphi_0$ & \si{\degree}                       & 0     & 0.2 \\
		Initial velocity          & $v_0$       & \si{\kilo\meter / \second}         & 7.2   & 0.036  \\
		Initial flight-path angle & $\gamma_0$  & \si{\degree}                       & -30   & 0.15 \\
		Initial heading angle     & $\chi_0$    & \si{\degree}                       & 45    & 0.225 \\
		Atmospheric correction    & $\xi$       &                                    & 1     & 0.05 \\
		\hline
		\hline
		Parameter                 & Symbol      & Unit                               & Value & \\
		\hline
		Initial altitude          & $h_0$       & \si{\kilo\meter}                   & 125   & \\
		Ballistic coefficient     & $\beta_0$   & \si{\kilo\gram / \square\meter} & 10000 & \\
		Lift coefficient          & $\alpha$    & \si{\square\meter / \kilo\gram} & 0     & \\
		\hline
	\end{tabular}
\end{table}

\begin{figure}[hbt!]
	\centering
	\includegraphics[width=0.85\textwidth]{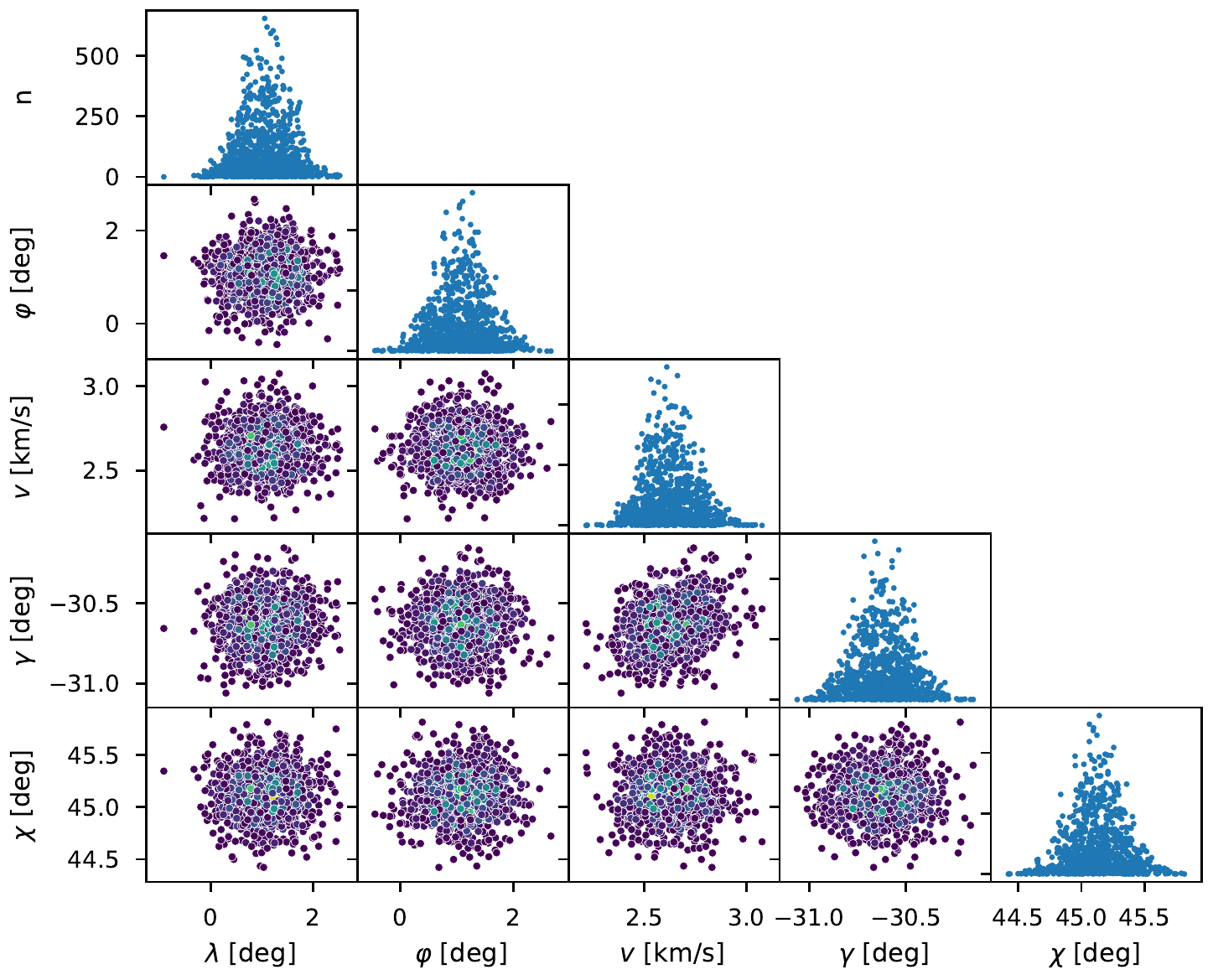}
	\caption{Landing snapshot for the six-state steep entry test case.}
	\label{fig:snapshot_earth6state}
\end{figure}

As it is possible to observe, in this case the uncertainties in the initial conditions have a smaller and more realistic standard deviation with respect to the ones presented in \cref{tab:strategic_state0}. The largest uncertainty is represented by the atmospheric correction coefficient, which introduces a 5\% standard deviation in the value of the atmospheric density. The uncertainty in the radial position has been removed as it is now the independent variable of the integration. This can allow setting a radius (altitude) value for the re-entry interface and exclude the radius from the uncertainties. The presented test case is intended to show the applicability of the continuum propagation to a realistic re-entry scenario, which considers the full dynamical representation, includes more realistic and accurate gravitational and atmospheric models, and takes into account the uncertainty in the atmospheric properties.
The six-dimensional distribution of the initial uncertainty is then sampled and each sample is propagated until the surface of the Earth is reached. \cref{fig:snapshot_earth6state} shows the projections of the distribution of 1000 sampled points at the end of the propagation (Earth impact). For the sake of readability, the distribution in the atmospheric correction coefficient $\xi$ is omitted; however, as it is a coefficient, its derivative is zero (\cref{eq:system6state}) and the distribution at the final instant is the same as the initial one. On the main diagonal we can observe the magnitude of the probability density associated to each sampled point.

It is interesting to observe how the distribution for this six-state variation of the \emph{strategic} test case is considerably more regular and maintains a more Gaussian-like behavior throughout the re-entry process when compared to its three-state counterpart (\cref{fig:strategic_scatter}). The difference is due to the different uncertainties in the initial conditions, which are much more pronounced for the test case in \cref{subsec:test_strategic}. The different independent variable (from $t$ to $r$) may also have contributed to the different shape of the state space.

\subsubsection{Comparison} \label{subsubsec:earth6state_results}

\begin{figure}[hbt!]
	\centering
	\begin{subfigure}[b]{0.32\textwidth}
		\includegraphics[height=4.6cm]{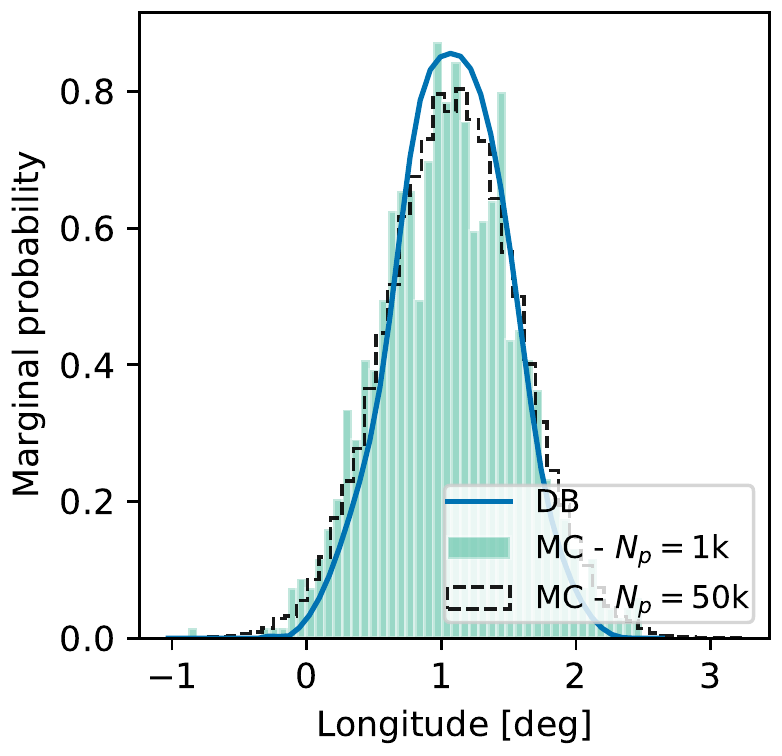}
		\caption{ }
		\label{fig:m1d_earth_lon}
	\end{subfigure}
	~
	\begin{subfigure}[b]{0.32\textwidth}
		\includegraphics[height=4.6cm]{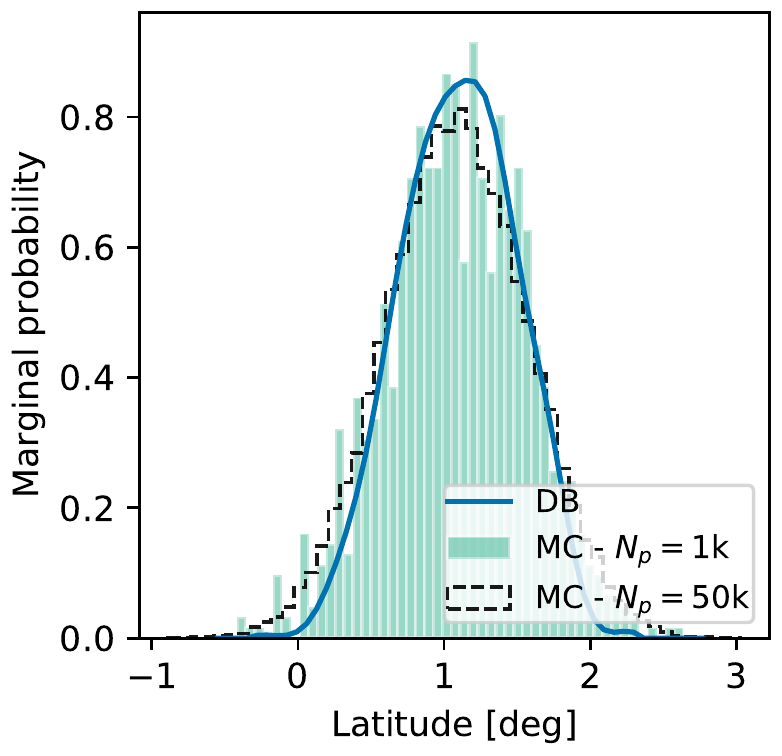}
		\caption{ }
		\label{fig:m1d_earth_lat}
	\end{subfigure}
	~
	\begin{subfigure}[b]{0.32\textwidth}
		\includegraphics[height=4.6cm]{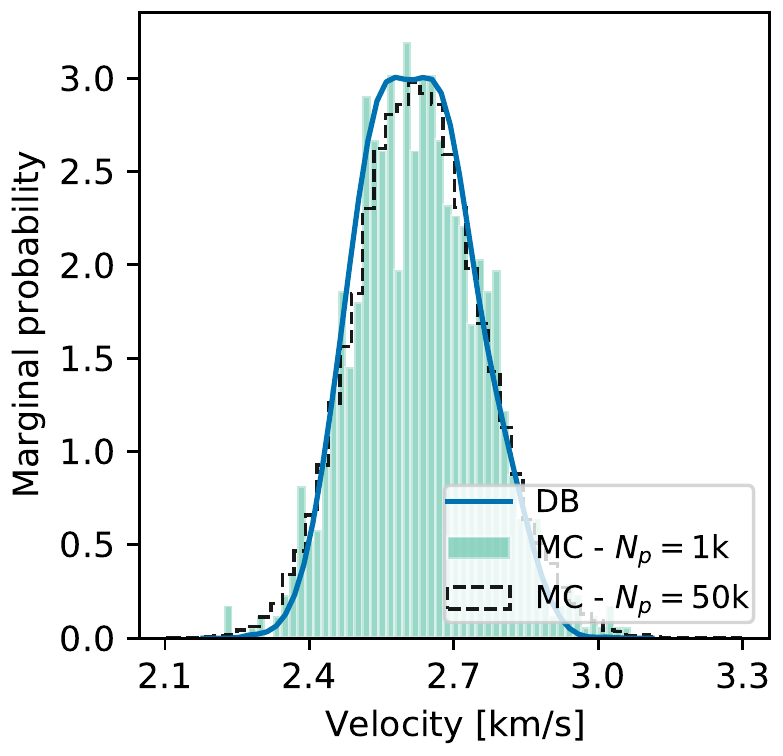}
		\caption{ }
		\label{fig:m1d_earth_v}
	\end{subfigure} \\
	~
	\begin{subfigure}[b]{0.32\textwidth}
		\includegraphics[height=4.6cm]{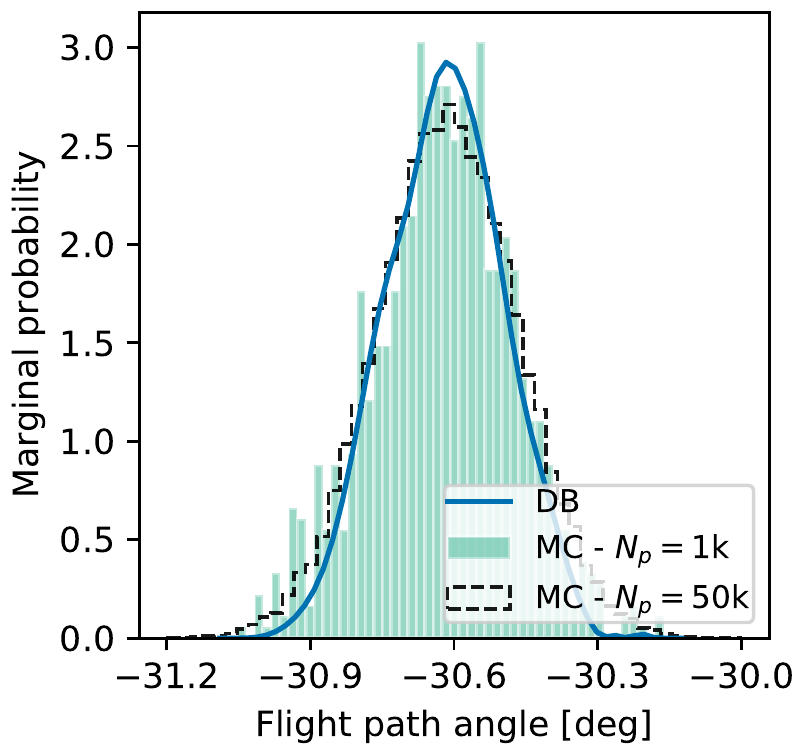}
		\caption{ }
		\label{fig:m1d_earth_fpa}
	\end{subfigure}
	~
	\begin{subfigure}[b]{0.32\textwidth}
		\includegraphics[height=4.6cm]{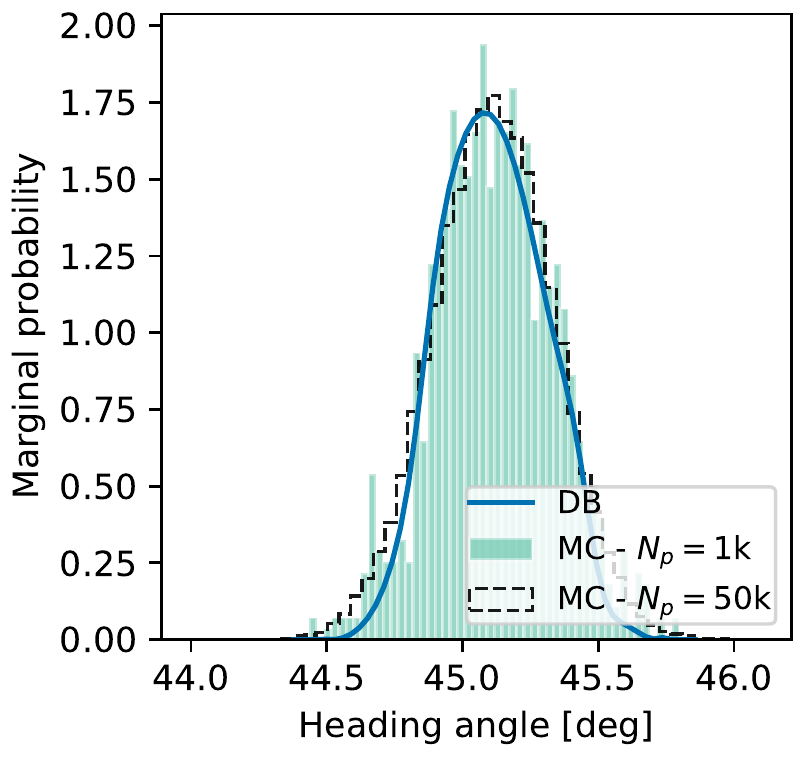}
		\caption{ }
		\label{fig:m1d_earth_head}
	\end{subfigure}
	
	\caption{DB vs. MC one-dimensional marginals comparison at landing instant for 1000 samples and 50000 samples.}
	\label{fig:m1d_earth6state}
\end{figure}

\cref{fig:m1d_earth6state} shows the comparison between the one-dimensional marginals obtained with the DB and the MC methods for the relevant state variables. The comparison is between the DB marginals obtained with 1000 samples, and the corresponding MC marginals obtained with 1000 (the green shaded histogram) and 50000 samples (the black dashed histogram) for the final propagation instant, which correspond to landing.

\begin{figure}[hbt!]
	\centering
	\begin{subfigure}[b]{0.45\textwidth}
		\includegraphics[height=4.6cm]{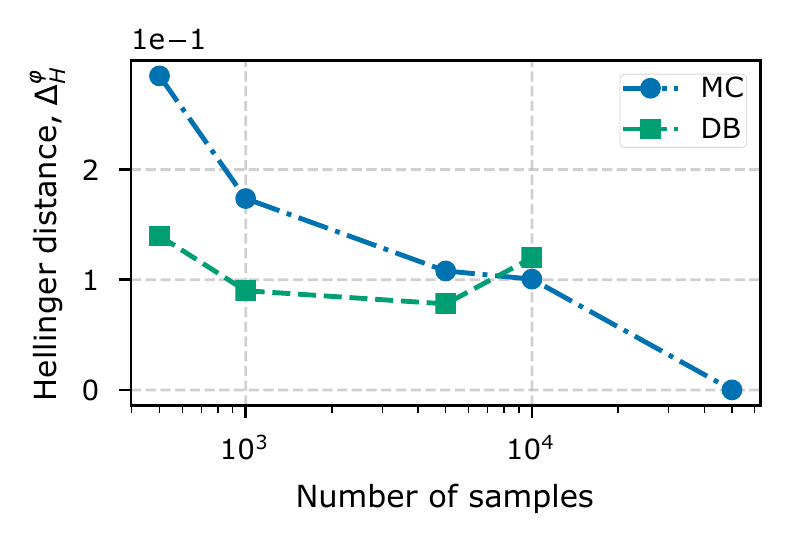}
		\caption{Hellinger distance for $\boldsymbol{\varphi}$}
		\label{fig:deltaH_6state_lat}
	\end{subfigure}
	~
	\begin{subfigure}[b]{0.45\textwidth}
		\includegraphics[height=4.6cm]{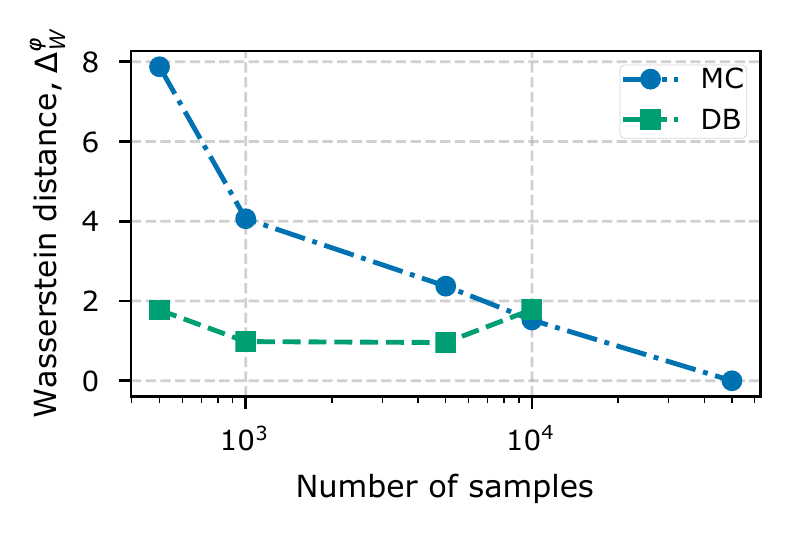}
		\caption{Wasserstein distance for $\boldsymbol{\varphi}$}
		\label{fig:deltaW_6state_lat}
	\end{subfigure} \\
	\begin{subfigure}[b]{0.45\textwidth}
		\includegraphics[height=4.6cm]{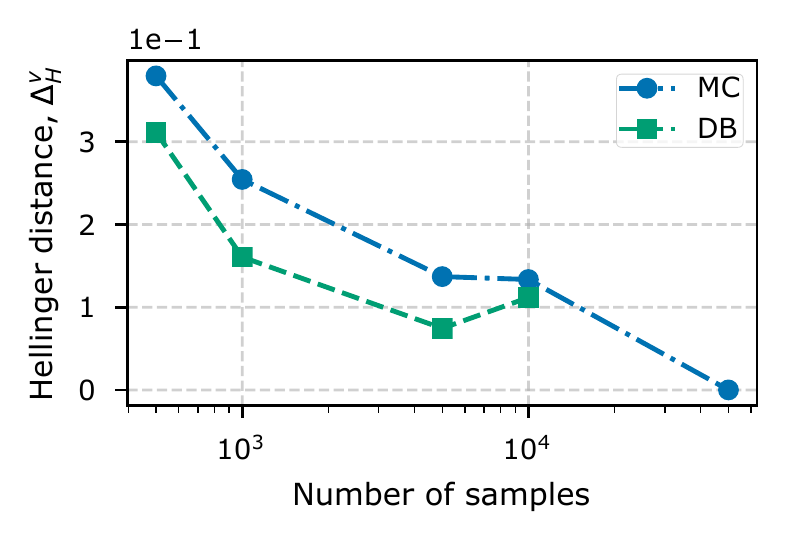}
		\caption{Hellinger distance for $\boldsymbol{v}$}
		\label{fig:deltaH_6state_v}
	\end{subfigure}
	~
	\begin{subfigure}[b]{0.45\textwidth}
		\includegraphics[height=4.6cm]{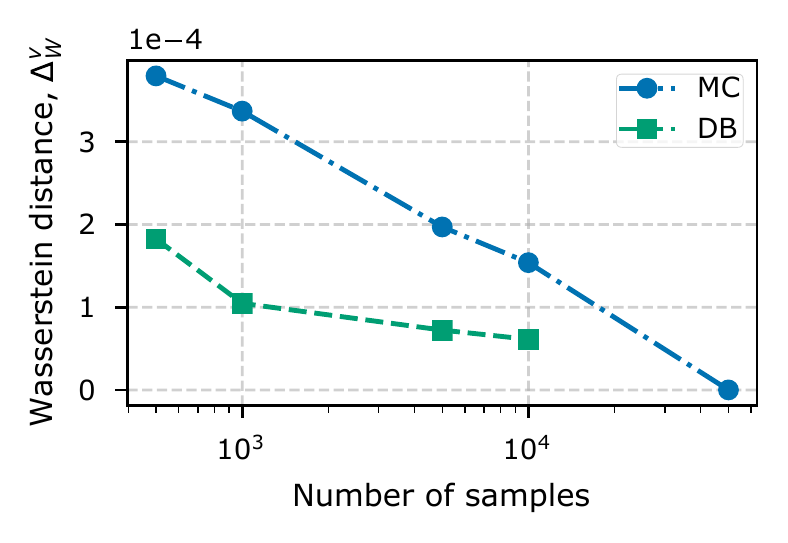}
		\caption{Wasserstein distance for $\boldsymbol{v}$}
		\label{fig:deltaW_6state_v}
	\end{subfigure} \\
	\begin{subfigure}[b]{0.45\textwidth}
		\includegraphics[height=4.6cm]{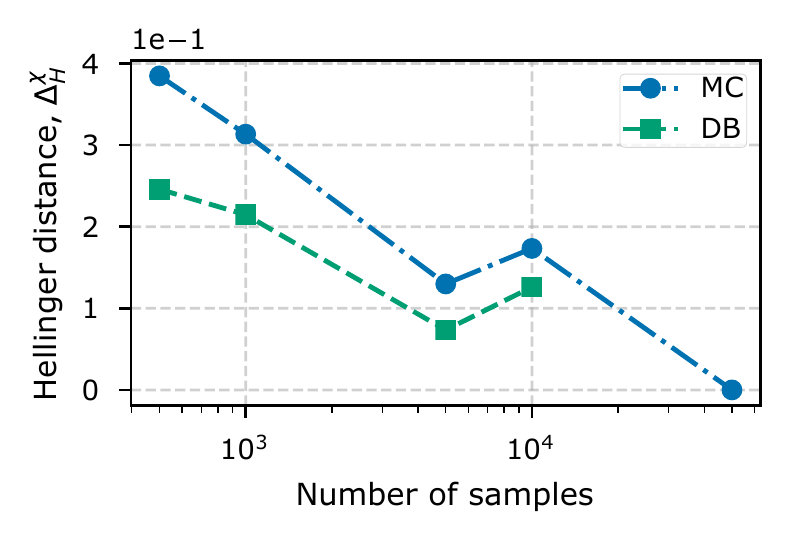}
		\caption{Hellinger distance for $\boldsymbol{\chi}$}
		\label{fig:deltaH_6state_head}
	\end{subfigure} 
	~
	\begin{subfigure}[b]{0.45\textwidth}
		\includegraphics[height=4.6cm]{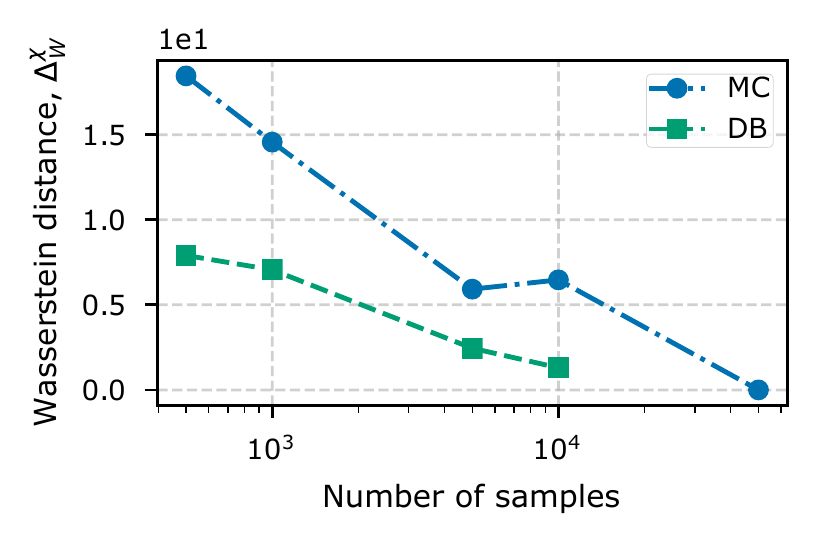}
		\caption{Wasserstein distance for $\boldsymbol{\chi}$}
		\label{fig:deltaW_6state_head}
	\end{subfigure} \\
	
	\caption{Variation of the Hellinger and Wasserstein distances between the DB and MC methods as a function of the number of samples for the landing snapshot.}
	\label{fig:comparison6state}
\end{figure}

The marginals show a similar behavior with respect to their three-state counterparts (\cref{fig:m1d_earth3state}): the distributions obtained with 1000 Monte Carlo samples show a more erratic behavior for all the states, especially in the areas around of the peak of the distribution, while the corresponding DB marginals show a more regular behavior. However, even in this case, it is possible to observe the feature of the DB marginals when compared to the 50000 Monte Carlo that is the presence of  peaks and smaller tails, which in turn results in a smaller standard deviation in the prediction of the states at the landing instant. To better compare the quality of the distribution reconstruction with both the DB and the MC methods, a comparison using the metrics introduced in \cref{subsec:test_strategic} (Hellinger distance and Wasserstein distance) is performed. \cref{fig:comparison6state} shows this comparison for three selected states (latitude, velocity, and heading angle). In the comparison the distances have been computed taking as reference distribution the Monte Carlo simulation with 50000 samples. In case of the DB methodology, the results for 50000 samples are not present as the memory requirements were too high for the laptop used in this study.

\begin{figure}[hbt!]
	\centering
	\includegraphics[width=0.45\textwidth]{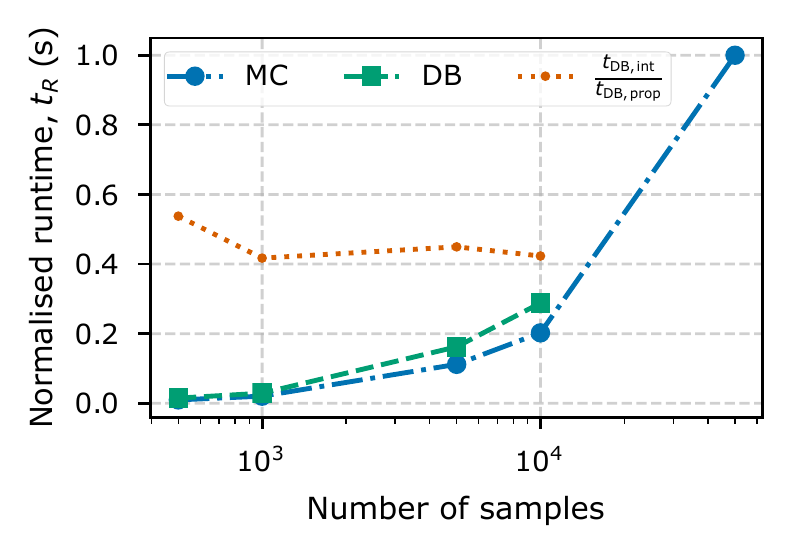}
	\caption{Run-time comparison between the DB and MC methodology. In orange, the fraction between the interpolation and the propagation times for the DB method.}
	\label{fig:time_6state}
\end{figure}

The results of the comparison show that the marginals obtained with the DB method consistently perform better with respect to their MC counterparts with the same number of samples according to both the proposed metrics. However, some notable features can be identified: the Hellinger distance shows in all three comparisons a worsening in the performance of the DB method when passing from 5000 to 10000 samples. This feature is confirmed also by the Wasserstein distance for the latitude marginal. The reason behind this behavior probably resides in the methodology used to obtain the marginal distribution as outlined in \cref{subsec:marginals} when combined with the increasing requirements for computational resources due to the growing number of points: in fact, to limit the memory usage a higher number of bins has been used, leading to a lesser performance in the estimation of the marginals. On the other hand, the Wasserstein distance (\cref{fig:deltaW_6state_lat,fig:deltaW_6state_v,fig:deltaW_6state_head}) shows that the DB marginals obtained with 1000 samples have a better or comparable result with respect to the 10000 MC simulation. A similar trend can be observed in two out of three comparisons of the Hellinger distance, where the heading angle case has a more pronounced difference between the 1000 DB and the 10000 MC simulations. Another interesting aspect, when comparing the results of \cref{fig:comparison6state} with the results of the previous test case \cref{fig:comparison_3state} is that in the last instant of the six-state case the DB method still performs better then the MC, while for the three-state case it has a worse performance. This feature is probably related to the fact the in this case, by introducing the radius as independent variable, all the samples are maintained in the snapshot until the end, while in the case of \cref{subsec:test_strategic} they are removed when they reach the ground, thus reducing the number of samples available for the fitting.
A final comparison is provided for the run-time of the code as a function of the number of points as shown in \cref{fig:time_6state}. The run-time has been normalized with respect to the run-time of the 50000 MC simulations so that all the other figures are a fraction of this reference time. Similarly to \cref{subsubsec:earth3state_results}, the run-times include the propagation time of the trajectory and the fitting time. As expected, for the same number of points, the DB method is always more time consuming as the fitting procedure requires more time than the histogram generation. However, it is also possible to observe that the DB method with 1000 samples requires about one-fourth and one-seventh of the time required by the 5000 and 10000 samples MC method respectively, with a performance that is comparable or worse than the one of the DB method, as shown in \cref{fig:comparison6state}. In addition, the \cref{fig:time_6state} shows the ratio between the marginalization time and the propagation time for the DB method, which oscillates between 40\% and 60\%.

% --------------------------------------------------------------
% TEST CASE - MARS 6 STATE
% --------------------------------------------------------------
\subsection{Six-state Mars re-entry}  \label{subsec:test_mars}
The third test case features a six-state Mars re-entry, whose propagation is again performed using \cref{eq:system6state}. The initial conditions and the other relevant re-entry parameters are based on the Mars MSL mission \citep{kluever2008entry,lunghi2018atmospheric} and they are summarized in \cref{tab:mars6state0}. The modified lift coefficient, $\alpha=0.001$ \si{\square\meter / \kilo\gram}, corresponds to a $C_L = 0.18$ of a spacecraft with mass $m=2200$ \si{\kilo\gram} and cross-section $S=12.9$ \si{\squared\meter} \citep{kluever2008entry}. 

\begin{table}[hbt!]
	\caption{\label{tab:mars6state0} Mean and standard deviation for the initial conditions of the augmented state space and values of the relevant parameters.}
	\centering
	\begin{tabular}{lcccc}
		\hline
		State                     & Symbol      & Unit                               & $\mu$ & $\sigma$ \\
		\hline
		Initial longitude         & $\lambda_0$ & \si{\degree}                       & -90.07      & 0.5 \\
		Initial latitude          & $\varphi_0$ & \si{\degree}                       & -43.9      & 0.5 \\
		Initial velocity          & $v_0$       & \si{\kilo\meter / \second}         & 5.505  & 0.004  \\
		Initial flight-path angle & $\gamma_0$  & \si{\degree}                       & -14.15 & 0.023 \\
		Ballistic coefficient     & $\beta$   & \si{\kilo\gram / \square\meter}    & 125.0  & 3.5 \\
		Atmospheric correction    & $\xi$       &                                    & 1      & 0.05 \\
		\hline
		\hline
		Parameter                 & Symbol      & Unit                               & Value & \\
		\hline
		Initial altitude             & $h_0$       & \si{\kilo\meter}                    & 126                   & \\
		Initial heading angle        & $\chi_0$    & \si{\degree}                        & 4.99                    & \\
		Lift coefficient             & $\alpha$    & \si{\square\meter / \kilo\gram}     & 0.001                     & \\
		Mars equatorial radius       & $R_p$       & \si{\kilo\meter}                    & 3397                  & \\
		Mars gravitational parameter & $\mu_p$     & \si{\cubic\meter / \second\squared} & 4.283$\times 10^{13}$ & \\
		Mars rotational rate         & $\omega_p$  & \si{\radian / \second}              & 7.095$\times10^{-5}$  & \\
		\hline
	\end{tabular}
\end{table}

\begin{figure}[hbt!]
	\centering
	\includegraphics[width=0.85\textwidth]{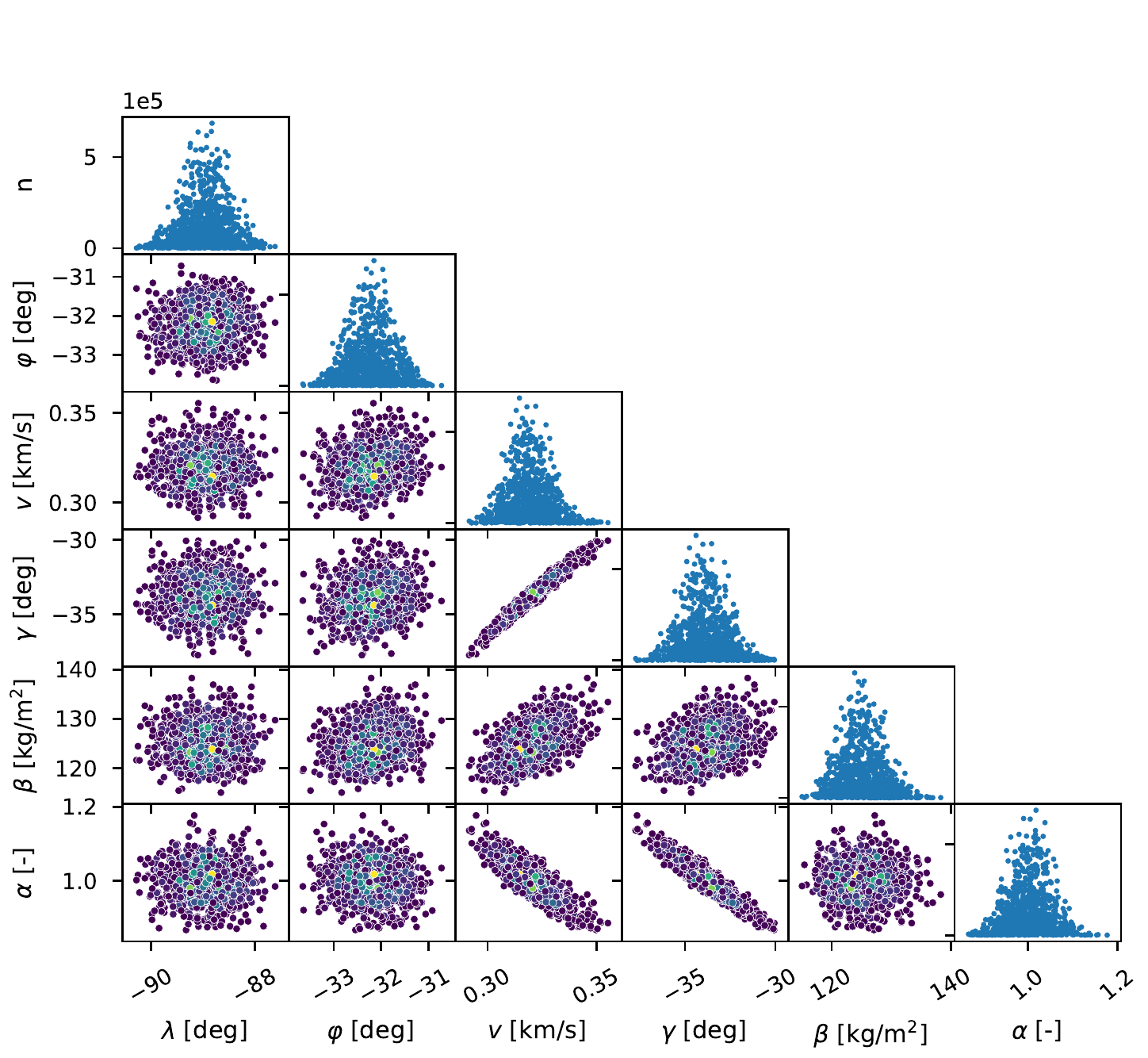}
	\caption{Pairplot of the landing snapshot for the six-state unguided Mars entry test case.}
	\label{fig:snapshot_mars6state}
\end{figure}

As it is possible to observe, in this case we have considered a lifting entry with Gaussian uncertainties in the initial longitude, latitude, velocity, and flight-path angle together with the ballistic coefficient and the atmospheric correction coefficient. The gravity model for Mars is an inverse exponential in this case, equivalent to the one adopted in \cref{subsec:test_strategic}. The atmospheric density is a function of the altitude has been obtained from the curve fitting of the Mars Science Laboratory (MSL) data and has the following expression \citep{lunghi2018atmospheric}:

\begin{equation}
	\rho(h) = \exp{c_0 + c_1 h + c_2 h^2 + c_3 h^3 + c_4 h^4},
\end{equation}

where $c_0 = -4.343$, $c_1 = -9.204\times10^{-5}$, $c_2 = -1.936\times10^{-11}$, $c_3 = -7.507\times10^{-15}$, and $c_4 = 4.195\times10^{-20}$. Again, the uncertainty in the atmospheric density is implemented following the procedure outlined by \cref{eq:atmCorrection}, introducing the atmospheric correction coefficient, $\xi$. \cref{fig:snapshot_mars6state} shows the projections of the snapshot at the landing instant for the Mars test case, where we have excluded the ballistic coefficient for a better readability. We can observe that, differently from the previous test case (\cref{fig:snapshot_earth6state}), the effect of the nonlinear dynamic is more pronounced and this translates into a more deformed state space, particularly in the coupling between the velocity, the flight-path angle and the atmospheric correction coefficient (the ballistic coefficient has a similar dependency).

\subsubsection{Comparison}  \label{subsubsec:mars_results}
\begin{figure}[hbt!]
	\centering
	\begin{subfigure}[b]{0.45\textwidth}
		\centering
		\includegraphics[height=5.2cm]{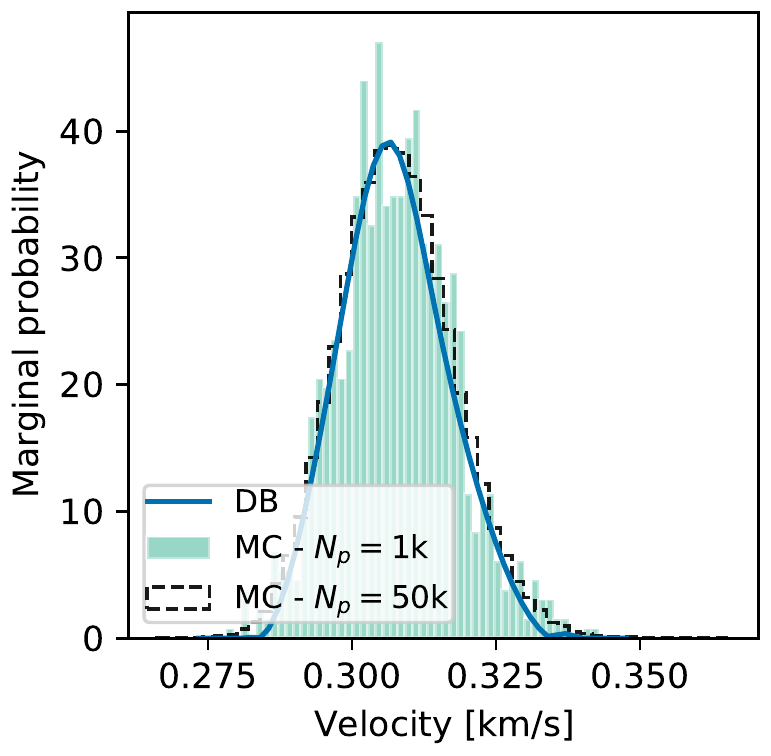}
		\caption{ }
		\label{fig:m1d_mars_v}
	\end{subfigure}
	~
	\begin{subfigure}[b]{0.45\textwidth}
		\centering
		\includegraphics[height=5.2cm]{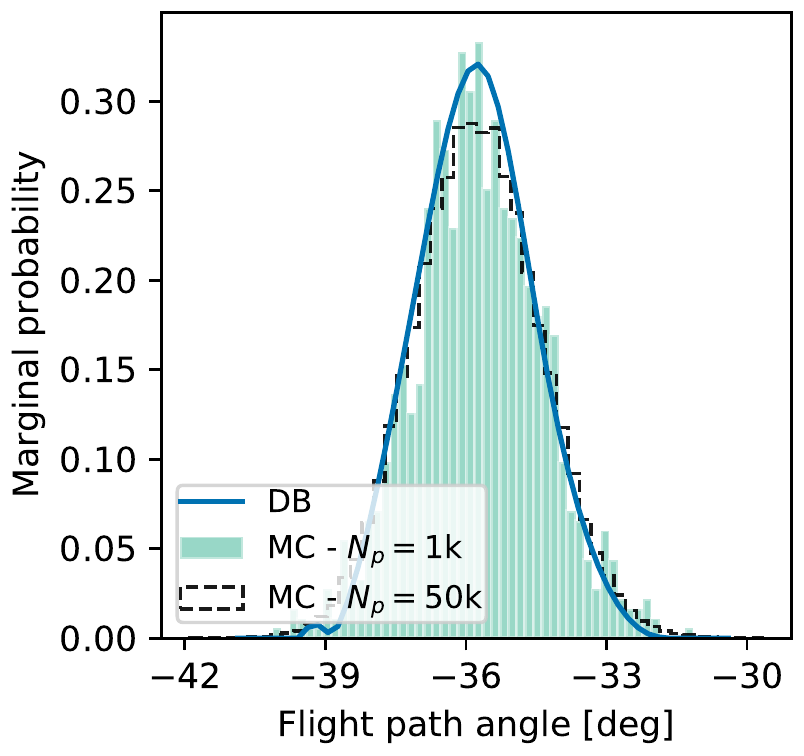}
		\caption{ }
		\label{fig:m1d_mars_fpa}
	\end{subfigure}
	
	\caption{DB vs. MC one-dimensional marginals comparison at landing instant for the Mars re-entry test case. DB method with 1000 samples and MC method with 1000 and 50000 samples.}
	\label{fig:m1d_mars6state}
\end{figure}

\begin{figure}[hbt!]
	\centering
	\begin{subfigure}[b]{0.32\textwidth}
		\centering
		\includegraphics[width=\textwidth]{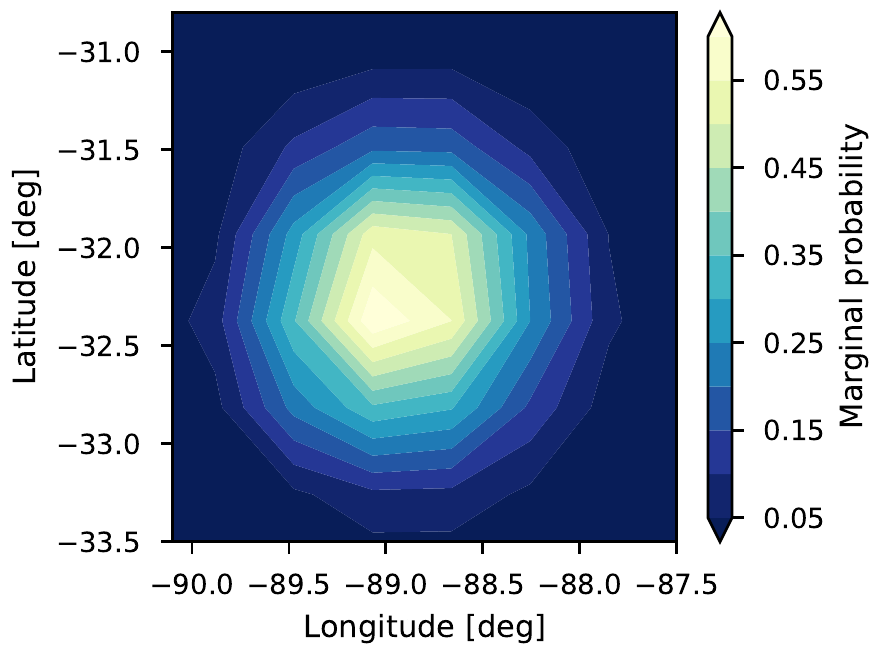}
		\caption{ }
		\label{fig:m2d_mars_db}
	\end{subfigure}
	~
	\begin{subfigure}[b]{0.32\textwidth}
		\centering
		\includegraphics[width=\textwidth]{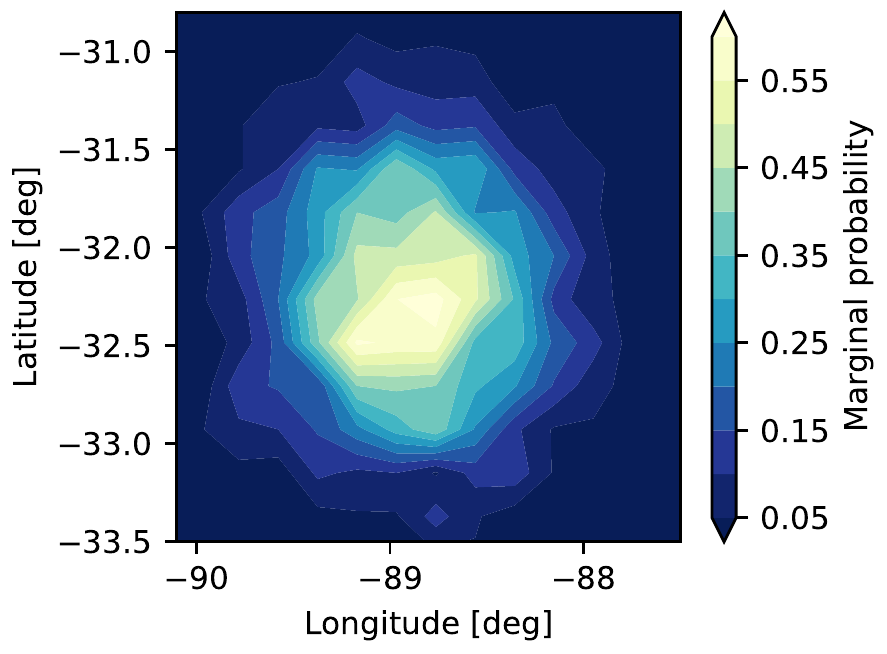}
		\caption{ }
		\label{fig:m2d_mars_mc_2k}
	\end{subfigure}
	~
	\begin{subfigure}[b]{0.32\textwidth}
		\centering
		\includegraphics[width=\textwidth]{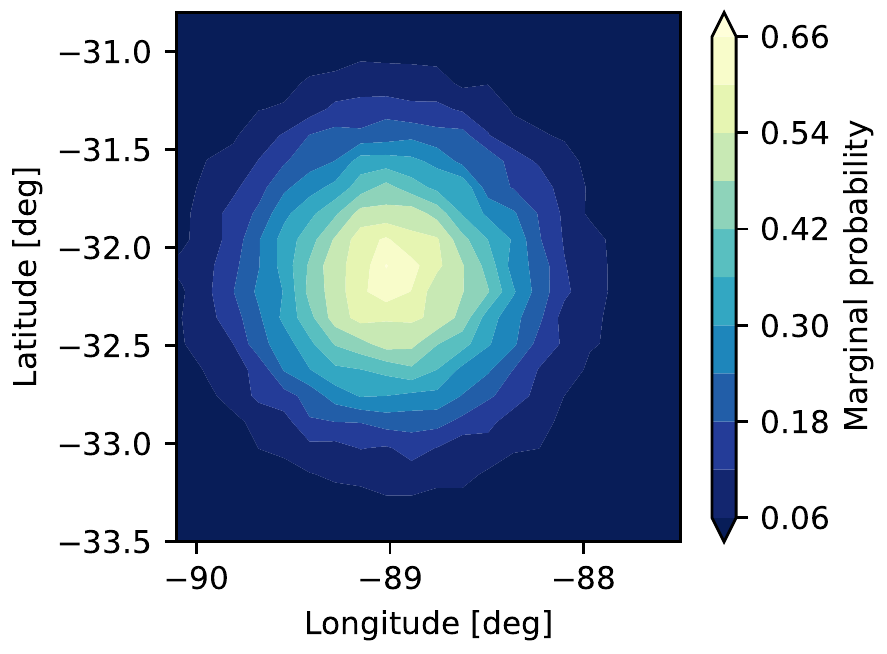}
		\caption{ }
		\label{fig:m2d_mars_mc_50k}
	\end{subfigure}
	
	\caption{DB vs. MC two-dimensional $\boldsymbol{\lambda}$-$\boldsymbol{\varphi}$ marginals comparison at landing instant for the Mars re-entry test case. DB method with 2000 samples (a) and MC method with 2000 (b) and 50000 samples (c).}
	\label{fig:m2d_mars6state}
\end{figure}

\cref{fig:m1d_mars6state} show the one-dimensional marginals obtained with the DB and MC method for the velocity and the flight-path angle at the landing instant. The DB marginal has been obtained with 1000 samples, while the MC method shows the results obtained with 1000 samples (green shaded histogram) and 50000 samples (black dashed histogram). The results show a good agreement of the DB method with the 50000 samples MC simulation taken as a reference. However, it is still visible the discrepancy around the tails of the distribution, with a more erratic behavior and a tendency to reproduce lower values of the marginal probability, as already seen in \cref{fig:m1d_earth6state,fig:m1d_earth3state}. \cref{fig:m2d_mars6state} instead shows the comparison between the two-dimensional marginals obtained with the DB and MC methods for the landing location coordinates expressed in longitude and latitude.

\begin{figure}[hbt!]
	\centering
	\begin{subfigure}[b]{0.45\textwidth}
		\centering
		\includegraphics[width=\textwidth]{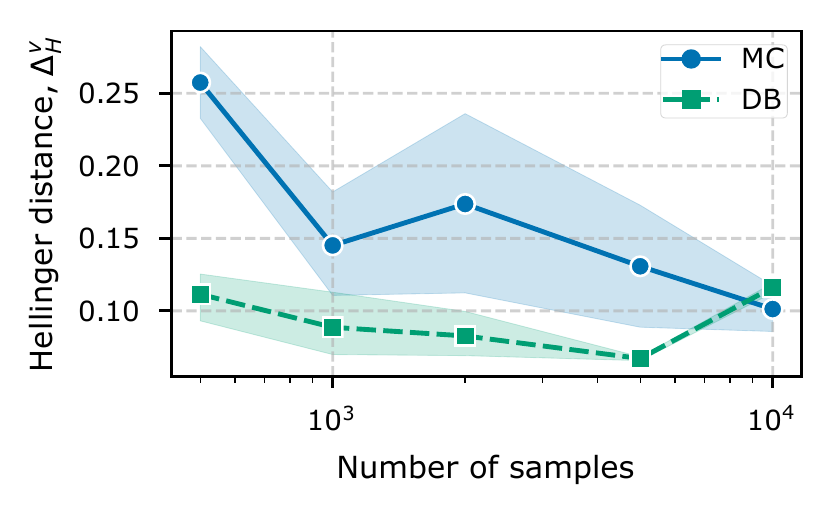}
		\caption{Hellinger distance for $\boldsymbol{v}$}
		\label{fig:m1d_hellinger_v}
	\end{subfigure}
	~
	\begin{subfigure}[b]{0.45\textwidth}
		\centering
		\includegraphics[width=\textwidth]{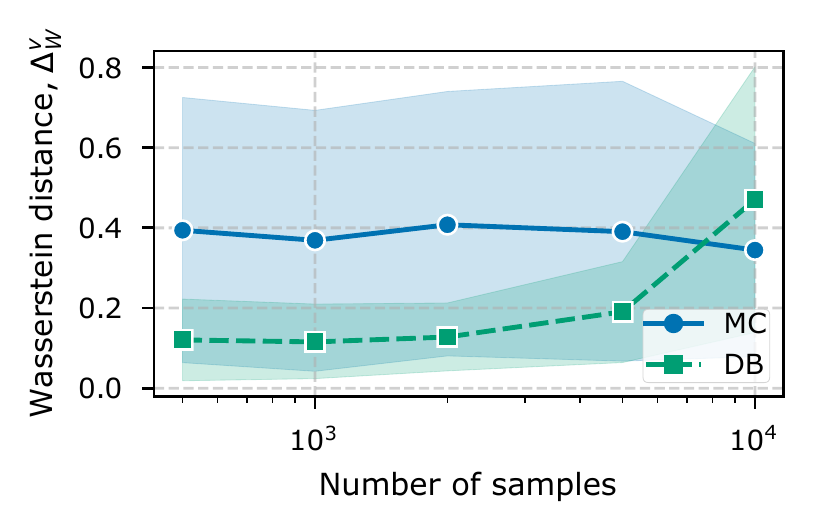}
		\caption{Wasserstein distance for $\boldsymbol{v}$}
		\label{fig:m1d_wasserstein_v}
	\end{subfigure} \\
	\begin{subfigure}[b]{0.45\textwidth}
		\centering
		\includegraphics[width=\textwidth]{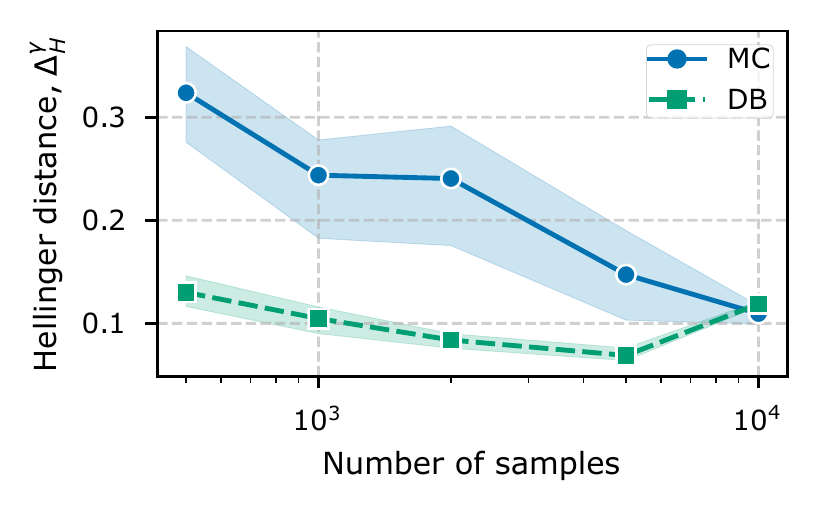}
		\caption{Hellinger distance for $\boldsymbol{\gamma}$}
		\label{fig:m1d_hellinger_fpa}
	\end{subfigure}
	~
	\begin{subfigure}[b]{0.45\textwidth}
		\centering
		\includegraphics[width=\textwidth]{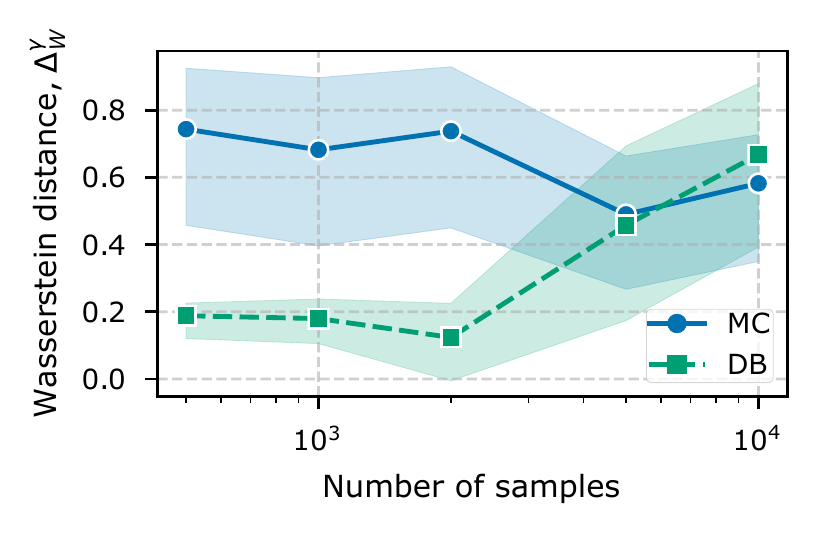}
		\caption{Wasserstein distance for $\boldsymbol{\gamma}$}
		\label{fig:m1d_wasserstein_fpa}
	\end{subfigure}
	\caption{Comparison of the average Hellinger and Wasserstein distances as a function of the number of samples for the variables $\boldsymbol{v}$ and $\boldsymbol{\gamma}$.}
	\label{fig:m1d_mars_comparison}
\end{figure}

A quantitative comparison between the obtained marginals is again performed exploiting the metrics introduced in \cref{subsec:test_strategic}, i.e. the Hellinger and Wasserstein distances. The comparison is performed computing the average and standard deviation of the Hellinger and Wasserstein distances over time for different number of sample points. The distances are computed taking as reference the results obtained with the MC simulation with 50000 samples. \cref{fig:m1d_mars_comparison} shows this comparison for the velocity and the flight-path angle as a function of the number of points. The solid lines represent the mean value and the shaded area the standard deviation. The Hellinger distance (\cref{fig:m1d_hellinger_v,fig:m1d_hellinger_fpa}) shows a regular behavior in both the considered variables for the DB method, with a constant improvement in performance when going from 500 to 5000 points. A worsening of the approximation happens instead for 10000 points, as it was the case for the Earth re-entry test case (\cref{fig:comparison6state}). In addition, it is possible to observe that, on average, the DB method shows a better performance with respect to the MC method, except for the 10000 samples case, and a smaller standard deviation. The Wasserstein distance (\cref{fig:m1d_wasserstein_v,fig:m1d_wasserstein_fpa}) shows a similar behavior up to 5000 samples, where the DB method performs better than the MC. However, after the 5000 samples, a worsening can be observed, especially in the case of the flight-path angle. It is also noticeable a considerably higher standard deviation for both the DB and MC methods. Contrary to the Hellinger distance, the Wesserstein distance is not a value ranging between zero and one, and its value changes through time alongside the change in magnitude of the probability density. For this reason, in this comparison the value at each snapshot of $\Delta_W$ has been normalized so that the average over time could be performed.

The results obtained for the two-dimensional marginals can also be compared using the Hellinger distance and are shown in \cref{tab:comparison2Dmars}. The results shows how the Hellinger distance for the DB method with 1000 samples is comparable with the ones of the MC method with 10000 samples, confirm the trend observed in the one-dimensional marginals and in previous test cases. A similar trend is also observed in the performance of the DB method, with an improvement passing from 1000 to 5000 samples, and then a worsening at 10000 samples.

\begin{table}[hbt!]
	\caption{\label{tab:comparison2Dmars} Comparison of Hellinger distance for the $\boldsymbol{\lambda}$-$\boldsymbol{\varphi}$ marginal at landing between DB and MC methods.}
	\centering
	\begin{tabular}{l|cccc}
		 & \multicolumn{4}{c}{Number of points} \\
		\hline
		Method & 1000 & 2000 & 5000 & 10000 \\
		\hline
		DB & 0.12518 & 0.10061 & 0.08712 &  0.0983 \\
		MC & 0.27565 & 0.25616 & 0.16781 & 0.11713 \\
		\hline
	\end{tabular}
\end{table}

\begin{figure}[hbt!]
	\centering
	\includegraphics[width=0.5\textwidth]{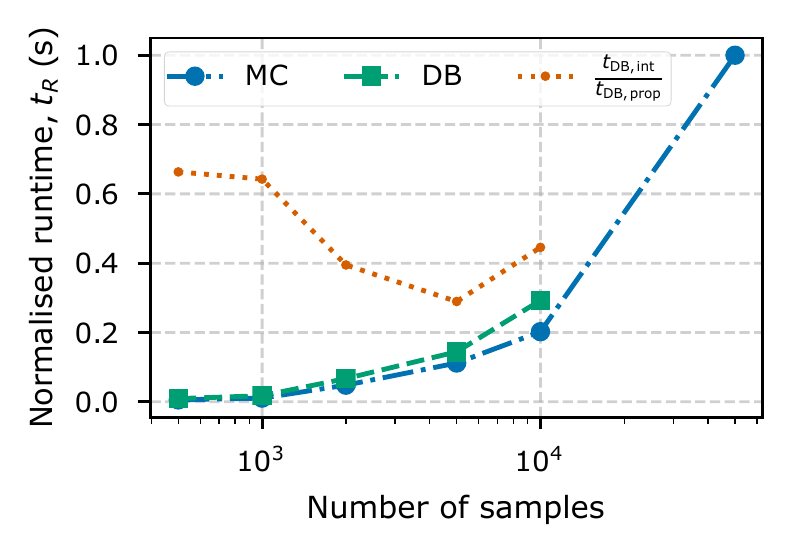}
	\caption{Run-time comparison between the DB and MC methodology. In orange, the fraction between the interpolation and the propagation times for the DB method.}
	\label{fig:time_6state_mars}
\end{figure}

Again, \cref{fig:time_6state_mars} provides a final comparison between the simulation times for the DB and MC methods. Also in this case the computational times have been normalized with respect to the 50000 MC simulation. As expected, the behavior closely matches the one of \cref{subsec:test_6stateEarth}, while the fraction of computational time devoted to the marginalization compared to the propagation shows a slightly less stable trend: higher fractions of the computational time is occupied by the marginalization procedure for smaller number of samples.

\subsubsection{Parachute deployment probability}  \label{subsubsec:parachute}
\begin{figure}[hbt!]
    \centering
    \includegraphics[width=0.45\textwidth]{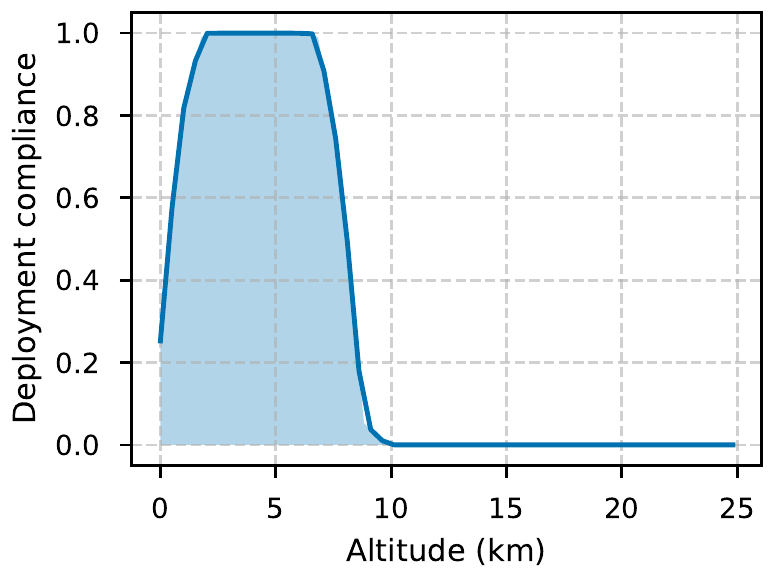}
    \caption{Parachute deployment compliance probability as a function of the altitude. Only the final part of the trajectory is shown.}
    \label{fig:parachute_prob}
\end{figure}

Similarly to the test case of \cref{subsec:test_strategic}, where we analyzed the probability of crossing the thresholds of dynamic pressure and heat rate, in this case, we consider a relevant problem in planetary entry (particularly on Mars) that is the deployment of a parachute to decelerate the probe in the final stage of the re-entry. For the parachute deployment, the main variables taken into account are the dynamic pressure and the Mach number. Particularly, the dynamic pressure is limited to $220 \leq \bar{q} \leq 880 \; \si{\newton / \square\meter}$, and the Mach number to $1.2 \leq M \leq 2.2$ \citep{kluever2008entry}. The dynamic pressure can be computed again using \cref{eq:pdyn}, while the Mach number has the following expression:

\begin{equation}  \label{eq:mach}
    M = \frac{v}{v_s},
\end{equation}

where $v_s$ is the speed of sound in martian atmosphere. Similarly to the atmospheric density profile, the speed of sound is considered only a function of the altitude and is obtained from a fitting of MSL data. The expression of the speed of sound as a function of the altitude is the following \citep{lunghi2018atmospheric}:

\begin{equation}  \label{eq:speed_of_sound}
	v_s(h) = 223.8 + c_1 h + c_2 h^2 + c_3 h^3,
\end{equation}

where $c_1 = -0.2\times10^{-3}$, $c_2 = -1.588\times10^{-8}$, and $c_3 = 1.404\times10^{-13}$. \cref{eq:pdyn,eq:mach} and the deployment boundaries in Mach number and dynamic pressure previously introduced can be combined to obtain the velocity limits within which both the limitations are satisfied. Using these boundaries with the one-dimensional marginals in velocity we can find the probability of being compliant with the parachute deployment limitations by integrating the velocity marginal within the boundaries. The procedure can be repeated for different snapshots to obtain the evolution of the compliance probability as a function of the altitude. \cref{fig:parachute_prob} shows the evolution of the compliance probability for the final portion of the trajectory. It can be noticed a full compliance within the altitude range 3 \si{\kilo\meter} and 7 \si{\kilo\meter}.

Mission planning and design for planetary entries can benefit from such analyses has they give a robust procedure for selecting the parachute deployment time given the state of the spacecraft under the effect of initial uncertainties. 

% ========================================================================
% CONCLUSION AND DISCUSSION
% ========================================================================

\section{Conclusion and discussion}  \label{sec:conclusion}
This work presents a novel methodology to propagate uncertainties using a continuum approach and a procedure to reconstruct these uncertainties and obtain the marginal probabilities of interest. The propagation methodology has been applied to relevant re-entry dynamics including both three-state and six-state dynamics and proved capable of including uncertainties in the state variables and in relevant parameters such as the ballistic coefficient and the atmospheric density. The procedure to include this uncertainties, however, requires an extension of the state space of the ODE so that the inclusion of a large number of uncertainty may result difficult to achieve at the current stage of development. The reconstruction methodology has been applied to representative test cases to derive the probability marginals for the quantities of interest. The test cases were representative of four-dimensional and six-dimensional state spaces and dynamics with different level of nonlinearity. The DB method showed the capability to obtain the marginal distribution up to a six-dimensional state space that also showed a substantial nonlinear behavior (\cref{fig:strategic_scatter,fig:snapshot_mars6state}). The results were compared with the corresponding Monte Carlo simulations for different number of samples, both in terms of difference between the obtained marginal distributions and simulation times. For the test case of \cref{subsec:test_strategic}, the DB method with 750 samples shows better performances in terms of the Hellinger and Wasserstein distances with respect to the MC method for the first part of the re-entry trajectory until about 28 seconds into the descent (\cref{fig:comparison_3state}). Instead, in the last part, the performance of the MC method slightly surpasses the one of the DB method, reaching a comparable performance at the final instant. For the test case of \cref{subsec:test_6stateEarth}, the DB method results in a better performance with respect to the MC simulations with an equivalent number of samples, except for the case with 10000 samples for the latitude variable (\cref{fig:deltaH_6state_lat,fig:deltaW_6state_lat}). The DB method also shows a similar or better Wasserstein distance with 1000 samples when compared with the MC method with 5000 and 10000 samples, while it has a worst performance when 750 samples are considered. The same trend can also be observed for the Hellinger distance for the latitude and velocity variables (\cref{fig:deltaH_6state_lat,fig:deltaH_6state_v}), but not for the heading angle (\cref{fig:deltaH_6state_head}). For the final test case of \cref{subsec:test_mars}, the DB method showed a better or comparable performance than the MC method in approximating the two-dimensional marginals as showed in \cref{tab:comparison2Dmars}. For the one-dimensional marginals, the DB method showed lower Hellinger distances when compared to the MC method with the same number of sample, except for the case with 10000 samples (\cref{fig:m1d_mars_comparison}). In addition, the DB method with as low as 750 samples shows, on average, a lower Hellinger distance than the MC method up to 5000 samples (\cref{fig:m1d_hellinger_v,fig:m1d_hellinger_fpa}). For the same comparison over a one-dimensional marginal, the Wasserstein distance shows a similar behaviour, with the DB method performing ,on average, better than the MC method for number of samples ranging from 750 to 5000 and instead losing the comparison for 10000 samples (\cref{fig:m1d_wasserstein_v,fig:m1d_wasserstein_fpa}). The computational time of the DB method is always greater than the MC when the same number of samples are used, because the marginalization procedure requires more computational time (\cref{fig:time_6state,fig:time_6state_mars}). However, this increase in time can, in some cases, be offset by using a lower number of samples to obtain a comparable level of approximation. The visual inspection of the obtained marginal probabilities showed a trend of the DB method to underpredict the tails of the distributions with respect to the MC method as confirmed by the results of \cref{fig:strategic_std_comparison}. This trend may be due to the lower number of samples available in these regions of the distributions, thus producing a less accurate reconstruction. In addition, another trend resulting from the presented analyses is the loss of performance of the DB method when a larger number of samples is used (i.e. 10000 samples in this work). This is a limitation of the introduced marginalization technique, which requires the definition of bins; if these bins are too small, the volume approximated by the triangulation may be inaccurate. However, a finer binning is usually required when the number of samples points increases (especially in higher dimensions) to manage the memory usage of the triangulation algorithm. This can limit the applicability of the reconstruction methodology to high-dimensional spaces. Still, the methodology has good performances when lower number of samples are used, and this is also where it is most applicable and appealing.

\section*{Funding Sources}
This project has received funding from the European Research Council (ERC) under the European Union’s Horizon 2020 research and innovation programme (grant agreement No 679086 - COMPASS). The data generated for this study can be found in the repository at the link www.compass.polimi.it/publications 

\bibliography{mybib}

\end{document}